\documentclass[12pt,preprint]{aastex}

\usepackage{amsmath,amssymb}
\usepackage{epsfig}
\usepackage{graphicx}
\usepackage{natbib}
\usepackage{subfigure}
\usepackage{mathrsfs}
\usepackage{float}
\usepackage{array}
\usepackage{rotating}
\usepackage{bm}

\newcommand{\myemail}{ywang12@hust.edu.cn}

\shorttitle{Coherent network analysis for continuous GW signals in PTA}

\shortauthors{Wang, Mohanty, & Jenet}

\begin{document}

\title{Coherent network analysis for continuous gravitational wave 
signals in a pulsar timing array: Pulsar phases as extrinsic parameters}

\author{Yan Wang\altaffilmark{1,2,3}, Soumya D. Mohanty\altaffilmark{2,4}, and Fredrick A. Jenet\altaffilmark{2,3,4}}

\affil{MOE Key Laboratory of Fundamental Physical Quantities Measurements,
 School of Physics, Huazhong University of Science and Technology, 1037 Luoyu Road, 
 Wuhan, Hubei Province 430074, China}\email{\myemail}
\affil{Department of Physics, University of Texas Rio Grande Valley, 
1 West University Boulevard, Brownsville, TX 78520, USA}
\affil{Center for Advanced Radio Astronomy, 1 West University Boulevard, Brownsville, TX 78520, USA}
\affil{Center for Gravitational Wave Astronomy,  1 West University Boulevard, Brownsville, TX 78520, USA}

\begin{abstract}

Supermassive black hole binaries are one of the primary targets for gravitational 
wave searches using pulsar timing arrays. Gravitational wave signals from such 
systems are well represented by parametrized models, allowing the standard 
Generalized Likelihood Ratio Test (GLRT) to be used for their detection and estimation. 
However, there is a dichotomy in how the GLRT can be implemented for pulsar 
timing arrays: there are two possible ways in which one can split the set of 
signal parameters for semi-analytical and numerical extremization. The 
straightforward extension of the method used for continuous signals in 
ground-based gravitational wave searches, where the so-called pulsar 
phase parameters are maximized numerically, was addressed in an earlier 
paper \citep{2014ApJ...795...96W}.
In this paper, we report the first study of the performance of the second 
approach where the pulsar phases are maximized semi-analytically. 
This approach is scalable since the number of parameters left over for numerical 
optimization does not depend on the size of the pulsar timing array. 
Our results show that, for the 
same array size (9 pulsars), the new method performs somewhat worse in parameter 
estimation, but not in detection, than the previous method where the pulsar phases 
were maximized numerically. The origin of the performance discrepancy is likely 
to be in the ill-posedness that is intrinsic to any network analysis method. 
However, scalability of the new method allows the ill-posedness to be mitigated by 
simply adding more pulsars to the array. This is shown explicitly by taking a 
larger array of pulsars.

\end{abstract}



\section{Introduction}

Pulsar timing array (PTA) based gravitational wave (GW) search is a promising 
approach for the very low frequency ($\sim10^{-9} -10^{-6}$ Hz) regime \citep{1978SvA....22...36S, 
1990ApJ...361..300F, 2005ApJ...625L.123J}, that is complimentary to the 
second-generation ground-based interferometers, such as Advanced LIGO \citep{2011arXiv1103.2728W}, 
Advanced Virgo \citep{2013ASPC..467..151D}, and KAGRA \citep{2012CQGra..29l4007S} 
operating at high frequencies ($\sim10 -10^3$ Hz), as well as to the space-based detectors, 
such as eLISA \citep{2013arXiv1305.5720C} proposed for low frequencies ($\sim10^{-4} -10^{-1}$ Hz). 
Unlike man-made instruments, PTA uses a network of high precision astronomical clocks, 
i.e., millisecond pulsars (MSPs), as a galactic-scale GW detector. Currently, three 
regional PTAs (NANOGrav\footnote{http://www.nanograv.org/}, 
PPTA\footnote{http://www.atnf.csiro.au/research/pulsar/ppta/} and 
EPTA\footnote{http://www.epta.eu.org/}) are operating at astrophysically 
interesting sensitivities that may lead to the detection of GWs in the near future. 
Shared data as well as collaborative and competitive efforts among individual PTAs 
bond them as the International Pulsar Timing Array (IPTA\footnote{http://www.ipta4gw.org/}, 
\citet{2013CQGra..30v4010M,2014arXiv1409.4579M}). The IPTA uses some of the most 
advanced radio telescopes in the world today to regularly monitor about 50 pulsars. 
Next generation radio telescopes with larger collecting areas and better backend 
systems, such as FAST \citep{2014arXiv1407.0435H} and SKA \citep{2009A&A...493.1161S}, 
will join the global observation campaign in the future  and push pulsar timing to 
higher precision and better detection sensitivities.

A promising GW signal for PTA is the stochastic 
background formed by the incoherent superposition of weak signals 
from a large unresolved population of supermassive black hole binaries (SMBHBs) 
\citep{1979ApJ...234.1100D, 1983ApJ...265L..35R, 
1990ApJ...361..300F, 2003ApJ...583..616J, 2003ApJ...590..691W,2005ApJ...625L.123J}. 
The stochastic GW perturbation will cause noise like signals in the pulsar 
time of arrivals (TOAs) that will be correlated across the pulsars in an array. 
The correlation will depend on the strength of the stochastic background and the 
pair-wise angular separation between the pulsars \citep{1983ApJ...265L..39H}.
Upper limits on the strength of the stochastic background along with our understanding 
of source  population have been improving over the years in correspondence 
with improvements in data quality \citep{2006ApJ...653.1571J, 2011MNRAS.414.1777Y, 
2011MNRAS.414.3117V, 2013Sci...342..334S, 2013ApJ...762...94D}.

In addition to the stochastic background, there exists the interesting possibility 
of detecting GWs from individual SMBHB sources 
\citep{1979ApJ...234.1100D, 2001ApJ...562..297L, 2004ApJ...606..799J, 2009MNRAS.400L..38S}. 
Simulations covering a range of massive black hole population models 
\citep{2009MNRAS.394.2255S,2010PhRvD..81j4008S} have shown that on average at least one source 
may be resolvable against the stochastic background. 
In the past few years, interest in 
analyzing continuous GW signals from individual SMBHBs has increased considerably 
\citep{2011MNRAS.414.3251L, 2011MNRAS.414...50D, 2012PhRvL.109h1104M, 2014arXiv1406.5297R}. 
Correspondingly, searches for continuous signals in the recent PTA data have been conducted in parallel with 
the stochastic background \citep{2010MNRAS.407..669Y, 2014arXiv1404.1267A, 2014MNRAS.444.3709Z}.

The detection and parameter estimation of continuous waves from individual sources in 
a PTA is a challenging data analysis task that has led to a number of studies 
\citep{2010MNRAS.407..669Y, 2010arXiv1008.1782C, 2012PhRvD..85d4034B, 
2012ApJ...756..175E, 2013CQGra..30v4004E, 2014ApJ...795...96W, 2014PhRvD..90j4028T, 
2015MNRAS.449.1650Z}. Unlike ground and space based detectors, the analysis of PTA 
data must contend with irregularly sampled time series with possible gaps, 
and noise components that must be estimated along with the signal as well 
as components that may be non-Gaussian or non-stationary \citep{2015inPrepYWANG}. 
As with any complex data analysis problem, a wide range of independent and 
complementary approaches are needed to build confidence in the final results.

This paper follows an earlier investigation reported in 
\citet{2014ApJ...795...96W} (hereafter WMJ1), where a Generalized 
Likelihood Ratio Test (GLRT) \citep{1998.book.....KayII} was constructed 
along the line of existing continuous wave signal searches used for 
ground based detectors  \citep{1998PhRvD..58f3001J, 2005PhRvD..72f3006C}. 
The WMJ1 method explicitly includes the \textit{pulsar terms} in the 
signal model and considers them as functions of pulsar phases. 
Numerical implementation of the GLRT usually involves a division of 
the signal parameters into the so-called extrinsic ones, over 
which the likelihood ratio can be maximized analytically or semi-analytically 
(including the use of Fast Fourier Transform), and the intrinsic ones for which a pure numerical 
optimization is required. However, unlike the ground based search, 
this division of the parameters into extrinsic and intrinsic is 
not unique in the case of a PTA. 
Following the convention used for the $\mathcal{F}$-statistic \citep{2012LRR....15....4J}, 
WMJ1 explored the choice that takes the overall amplitude of 
the signal, the inclination angle between the binary orbit and the plane of the sky, and 
the polarization angle of the GW as extrinsic and treated the pulsar phases and
remaining parameters as intrinsic. Our results showed that the pulsar phases 
are uninformative parameters, indicating that they are best 
marginalized or maximized as extrinsic parameters. 
The more important motivation to do so, however, is the fact that 
the number of pulsar phase parameters increases with the size of 
the PTA. Hence, the numerical optimization task will become infeasible 
at some point. Thus, the approach of treating pulsar phase parameters 
as intrinsic is not a scalable one.

This paper presents the first implementation of a method based on treating 
the pulsar phases as extrinsic parameters in a GLRT. Although, the idea of 
semi-analytical maximization over pulsar phases was presented in 
\citet{2012ApJ...756..175E}, 
a concrete implementation and performance characterization of the resulting 
method has not been reported until now. The method proposed here retains the 
use of Particle Swarm Optimization (PSO) to handle the numerical optimization 
over intrinsic parameters, but the particular variant of the PSO meta-heuristic 
used in this paper is different. 

An alternative to maximization over the pulsar phases is to marginalize 
over them following a Bayesian framework. This is the approach that has 
been studied the most in the PTA literature so far 
\citep{2013CQGra..30v4004E, 2014PhRvD..90j4028T}. To enable meaningful comparisons, 
the performance of the method presented here is studied on simulated data corresponding to
a PTA configuration adopted in \citet{2014PhRvD..90j4028T}. We use 
signal strengths, measured in terms of the network signal-to-noise ratio (S/N) $\rho_n$, that 
span a wide range from strong ($\rho_n=100$) to moderate ($\rho_n=30$) and 
barely detectable ($\rho_n=8$).  Although useful for testing the performance of the 
algorithm, $\rho_n>20$ is  unrealistic for PTA based GW detection in the foreseeable future. 
Thus the performance of the method for $\rho_n = 8$ to $\rho_n = 30$ serves to 
bracket the scenario that is more likely. 
As in WMJ1,  we simulate a large number of independent data realizations and
derive conventional Frequentist error estimates for the signal parameters. 

The results show that this method performs marginally better than the 
method in WMJ1 for detection, but the estimation of the angular parameters is 
somewhat worse. Specifically, while the localization of sources in WMJ1 
and the Bayesian method are comparable, shifting to a different split 
of extrinsic and intrinsic parameters creates secondary maxima that increase 
the scatter of estimated source locations.  This is most likely the result of 
the well known ill-posedness of the GW network analysis 
problem \citep{2005PhRvD..72l2002K,2006CQGra..23S.673R,2006CQGra..23.4799M}. 
Ill-posedness in inverse problems, such as GW network analysis, is marked 
by instability or discontinuity of the inferred solution under small perturbations 
in the data.  The source of perturbation can be either the noise in the data or 
numerical errors from computations. The jumping of solutions to radically 
different values can manifest itself as a large bias or large variance in estimation. 
For strictly linear models, such as GW burst searches where the time samples of the
two polarization waveforms directly form the parameters to be 
estimated \citep{2006CQGra..23S.673R}, ill-posedness is easily seen to be rooted 
in the rank deficiency of the matrix  $A^{T}A$, where $A$ is the 
$m\times 2$ network response matrix ($m$ is the number of detectors). 
The origin of ill-posedness in parameter estimation presented in this work 
is not as straightforward because the signal model is nonlinear in the 
parameters. 

Mitigation of ill-posedness requires regularization in some form, 
such as the imposition of constraints on the GLRT solutions 
\citep{1959SIAMR...1...38G,1977..book.....T}.  While some constraints appear 
naturally in the implementation of GLRT in WMJ1, they are absent in the 
formulation of the method presented here. 
The effects of ill-posedness are reduced, in general, by increasing the 
number of differently oriented detectors in a network. We demonstrate this by considering 
the case of a PTA with 17 pulsars. For this reason, we do not go deeper 
into the issue of regularization for PTA in this paper but leave it for 
future work to address.

The rest of the paper is organized as follows. In Section~\ref{sec:datamodel} 
we introduce the data model used in this paper.  Section~\ref{sec:MaxLR} 
describes the GLRT for this data model and its implementation, which involves 
maximization over pulsar phases analytically by solving quartic equations. 
Section~\ref{sec:applications} characterizes the method using simulated data 
and compares its performance with WMJ1 and \citet{2014PhRvD..90j4028T}.  
The paper is concluded in Section~\ref{sec:summary}. Some details about solving 
the quartic equation have been relegated to Appendix A.

\section{Data model}\label{sec:datamodel}

The data used for GW signal detection and parameter estimation in the case of 
a PTA consists of a set 
of timing residuals $r^I = (r^I_1,r^I_2,\ldots,r^I_{N_I})$, $I=1,2,\ldots,N_p$, 
where $N_p$ is the number of pulsars, $N_I$ is the number of observation for the 
$I$-th pulsar. Each timing residual is associated with a time of observation  
$t^I_i\in [0,T]$, $t^I_{i+1}>t^I_i$. The time interval between two 
observations can vary typically from several days up to a few weeks. When 
there is a signal in the data, $r^I_k = s^I_k+n^I_k$; otherwise, $r^I_k=n^I_k$. 
Here $n^I=(n^I_1,n^I_2,\ldots,n^I_{n_{I}})$ and $s^I = (s^I_1,s^I_2,\ldots,s^I_{n_{I}})$ 
denote the noise realization and the GW signal respectively.
The models for the signal and the noise (zero mean stationary Gaussian) remain 
the same as in WMJ1, but it is convenient to express 
the signal in a functional form that allows the pulsar phases to be easily 
extracted as extrinsic parameters in the detection statistic. 

GWs perturb the proper distance between a pulsar and an observer on the Earth, 
causing fluctuations of the TOAs of radio pulses with time. In the TT-gauge 
associated with a plane GW, the perturbation in the metric tensor can be written as 
\begin{equation}
\mathbf{h}=(h_+ \mathbf{e}_+ + h_{\times} \mathbf{e}_{\times}) e^{i(\omega_{\text{gw}}t-\mathbf{k}\cdot\mathbf{x})} \,,  
\end{equation}\label{gw}
where $\omega_{\text{gw}}$ is the GW angular frequency, $\mathbf{k}$ is the 
GW wave vector, and
\begin{subequations}\label{basepc}
\begin{align}
\mathbf{e}_+ = \boldsymbol{\hat{\alpha}}\otimes\boldsymbol{\hat{\alpha}} 
-\boldsymbol{\hat{\delta}}\otimes\boldsymbol{\hat{\delta}}  \label{first}   \,, \\
\mathbf{e}_\times = \boldsymbol{\hat{\alpha}}\otimes\boldsymbol{\hat{\delta}} 
+\boldsymbol{\hat{\delta}}\otimes\boldsymbol{\hat{\alpha}}  \label{second}   \,.
\end{align}
\end{subequations}
$\boldsymbol{\hat{\alpha}}$ and $\boldsymbol{\hat{\delta}}$ are the unit 
vectors along right ascension and declination in equatorial coordinates. 
The response of the detector to the GW is given by 
\begin{equation}
s^{I}_{i}(\lambda) = F^{I}_{+}(\alpha,\delta) \Delta h_{+}(t^{I}_{i};\theta) 
+ F^{I}_{\times}(\alpha,\delta) \Delta h_{\times}(t^{I}_{i};\theta)  \,,  
\end{equation}\label{resp}
where $F^{I}_{+}$ and $F^{I}_{\times}$ are the \textit{antenna pattern functions} 
(defined in Equations 9 and 10 of WMJ1), $\alpha$ and 
$\delta$ are the right ascension and declination of the source, $\theta$ represents 
collectively the following parameters: 
(i) $\zeta$, the overall amplitude factor (defined in Equation 7 of WMJ1); (ii) $\iota$, the 
inclination angle between the binary orbital plane and the plane of the sky; (iii) $\psi$, 
the GW polarization angle; (iv) $\varphi_0$, the initial phase of the binary at the beginning of 
the observation; (v) parameter $\varphi_I=\varphi_0-\frac{1}{2}\omega_{\text{gw}}d^I(1-\cos\theta^I)$, 
the pulsar phase parameter that contains the distance $d^I$ from the pulsar 
to Earth and the open angle $\theta^I$ between the lines of sight to the pulsar and 
the GW source. Hereafter, we regard the pulsar phases as independent variables. 
$\lambda=\lbrace\alpha,\delta\rbrace\cup\theta$ denotes the set of all the parameters. 
The term $\Delta h_{+,\times}(t^{I}_{i};\theta)$ is the difference of the 
GW tensor at Earth and at the pulsar at the observer's time $t^{I}_{i}$, 
\begin{equation}
\Delta h_{+,\times}(t^{I}_{i};\theta) =  h_{+,\times}(t^{I}_{i};\theta) 
- h_{+,\times}(t^{I}_{i} - \tau^{I};\theta) \,,  
\end{equation}\label{delhpc}
where $\tau^I=d^I(1-\cos\theta^I)/c$ is the time delay of the plane GWs of the same 
phase arriving at Earth and at the pulsar.  Hereafter, we assume that the binary 
system is evolving slowly, so that in the signal model the orbital frequency in the pulsar term 
remains approximately the same as in the Earth term.

The GW signal can be written as
\begin{eqnarray}\label{eq:residual2}
s^{I}_{i} &=& 2\zeta(1+\cos^2\iota)(F_+^{I}\cos 2\psi-F_\times^{I}\sin 2\psi)\sin(\varphi_0-\varphi_I )\sin(\varphi_0+\varphi_I+\Phi(t^{I}_{i}) )  \nonumber \\
                 &-& 4\zeta\cos\iota(F_+^{I}\sin 2\psi+F_\times^{I}\cos 2\psi)\sin(\varphi_0-\varphi_I )\cos(\varphi_0+\varphi_I+\Phi(t^{I}_{i}) ) \nonumber \\
                 &=& \mathcal{A}_I\sin(\varphi_0-\varphi_I) \sin(\varphi_0+\varphi_I+\phi_{I}+\Phi(t^{I}_{i})) \,,
\end{eqnarray}
where  $\Phi(t^I_i)= \omega_{\rm gw} t_i^I $, 
\begin{eqnarray}\label{eq:Ai}
\mathcal{A}_I^2 &=& 4\zeta^2(1+\cos^2\iota)^2(F_+^{I}\cos 2\psi-F_\times^{I}\sin 2\psi)^2 \nonumber \\
&+&16\zeta^2\cos^2\iota(F_+^{I}\sin 2\psi + F_\times^{I}\cos 2\psi)^2  \,,
\end{eqnarray}
and
\begin{equation}\label{eq:Tani}
\tan\phi_I=\frac{-2\cos\iota}{1+\cos^2\iota}\cdot\frac{F_+^{I}\sin 2\psi+F_\times^{I}\cos 2\psi}{F_+^{I}\cos 2\psi
-F_\times^{I}\sin 2\psi}  \,.
\end{equation}
Here $\mathcal{A}_I$ and $\phi_I$ depend only on $\zeta, \iota, \psi, \alpha, \delta$. 
In Equation~\ref{eq:residual2}, we can isolate the $\varphi_I$ dependence and get 
\begin{equation}\label{eq:residualphi}
s^{I}_{i}=(\mathcal{B}_I-\mathcal{E}_I)\cos 2\varphi_{I}+(\mathcal{C}_I
+\mathcal{D}_I)\sin2\varphi_{I}+(\mathcal{B}_I+\mathcal{E}_I) \,,
\end{equation}
where 
\begin{equation}\label{eq:Bi}
\mathcal{B}_{I}(t^{I}_i)= \frac{1}{2}\mathcal{A}_I\sin\varphi_0\sin(\varphi_0+\phi_{I}+\Phi(t^{I}_i)) \,,
\end{equation}
\begin{equation}\label{eq:Ci}
\mathcal{C}_{I}(t^{I}_i)= -\frac{1}{2}\mathcal{A}_I\cos\varphi_0\sin(\varphi_0+\phi_{I}+\Phi(t^{I}_i)) \,,
\end{equation}
\begin{equation}\label{eq:Di}
\mathcal{D}_{I}(t^{I}_i)= \frac{1}{2}\mathcal{A}_I\sin\varphi_0\cos(\varphi_0+\phi_{I}+\Phi(t^{I}_i)) \,,
\end{equation}
\begin{equation}\label{eq:Ei}
\mathcal{E}_{I}(t^{I}_i)= -\frac{1}{2}\mathcal{A}_I\cos\varphi_0\cos(\varphi_0+\phi_{I}+\Phi(t^{I}_i)) \,.
\end{equation}
Note that $\mathcal{B}_{I}$, $\mathcal{C}_{I}$, $\mathcal{D}_{I}$, $\mathcal{E}_{I}$ 
are functions of time and the source parameters rather than $\varphi_I$.

\section{Generalized Likelihood Ratio Test} \label{sec:MaxLR}

In the Frequentist approach, the detection of GW signals presents a composite 
hypotheses test problem: Given data ${\bf r}$, we need to pick one among a family 
of hypotheses about the joint probability density function (pdf) from which 
${\bf r}$ is obtained. Under the null hypothesis $\mathcal{H}_0$, the data 
does not contain any GW signal and the pdf, $p({\bf r})$, governing 
${\bf r}$ is that of the noise alone. Under the alternative hypothesis $H_\lambda$, 
a GW signal ${\bf s}(\lambda)$ with parameters $\lambda$ is present in ${\bf r}$ 
and the data is a realization from a governing pdf of the form 
$p({\bf r}|\lambda) = p({\bf r}-{\bf s}(\lambda))$. In a GLRT, assuming that 
the PDF of the noise $p(\mathbf{r})$ is known, the test statistic
\begin{equation}\label{eq:glrt}
\text{GLRT}(\mathbf{r})=\max_{\lambda}\frac{p(\mathbf{r}|\lambda)}{p(\mathbf{r})} 
=\max_{\lambda} \text{LR}(\mathbf{r};\lambda) = \text{LR}(\mathbf{r};\widehat{\lambda})\,,
\end{equation}
is compared with a threshold to decide in favor of $\mathcal{H}_0$ or 
$\mathcal{H}_{\widehat{\lambda}}$. Here, $\text{LR}(\mathbf{r};\lambda)$ 
is the likelihood ratio for a given hypothesis and $\widehat{\lambda}$ 
is the Maximum Likelihood Estimate (MLE) of the parameters. The maximizer, 
$\widehat{\lambda}$, of $\text{LR}(\mathbf{r};\lambda)$ in Eq.~\ref{eq:glrt} 
is the same as that of any monotonic function of $\text{LR}(\mathbf{r};\lambda)$. 
Its logarithm, $\Lambda(\mathbf{r};\lambda)$, is one such convenient choice 
for the case of Gaussian noise. 

Unlike the case of a known $\lambda$, where the optimal test statistic (under the
Neyman-Pearson criterion) is known to be $\text{LR}({\bf r};\lambda)$, 
there is no proof of optimality associated with 
the GLRT except in some simple cases. 
However, it has been shown that it is the 
\textit{uniformly most powerful} (UMP) 
among all \textit{invariant} tests \citep{1959tsh..book.....L}. In practice, 
and when it is computationally feasible, the GLRT is often found to be 
superior to other {\em ad hoc} tests.

\subsection{The network likelihood ratio} \label{sec:llr}

For a PTA of $N_p$ pulsars, the log-likelihood ratio is
\begin{eqnarray}
\Lambda(\mathbf{r};\lambda) & = & \sum_{I=1}^{N_p} \Lambda_I({\bf r};\lambda) \nonumber\;,\\
\Lambda_I({\bf r};\lambda) & = & \langle r^I|s^{I}(\lambda)\rangle_I
-\frac{1}{2}\langle s^{I}(\lambda)|s^{I}(\lambda)\rangle_I  \,,\label{eq:loglambda}
\end{eqnarray}
where $\langle\cdot|\cdot\rangle_I$ is the \textit{noise weighted inner product}, 
$(\cdot){\bf C}_I^{-1}(\cdot)^T$, with ${\bf C}_I$ being the covariance matrix 
of the noise process in the $I$-th pulsar. It is assumed here that the 
cross-covariances of noise between $r^I$ and $r^J$ are ignorable for 
$I \neq J$. Inserting Eq.~\ref{eq:residualphi} into Eq.~\ref{eq:loglambda} 
we have 
\begin{align}
\Lambda_I(\mathbf{r};\lambda) & = \left[ \langle r^I|X_{I}\rangle_I\cos2\varphi_{I}
+\langle r^I|Y_{I}\rangle_I\sin2\varphi_{I}+\langle r^I|Z_{I}\rangle_I \right. \nonumber \\
& \left. -\frac{1}{2} \left(\langle X_{I}|X_{I}\rangle_I\cos^{2} 2\varphi_{I}
+\langle Y_{I}|Y_{I}\rangle_I\sin^{2} 2\varphi_{I}+2\langle X_{I}|Y_{I}\rangle_I\sin2\varphi_{I}\cos2\varphi_{I} \right. \right. \nonumber  \\
& +2\langle X_{I}|Z_{I}\rangle_I\cos2\varphi_{I}+2\langle Y_{I}|Z_{I}\rangle_I\sin2\varphi_{I}
+\langle Z_{I}|Z_{I}\rangle_I \left.\left. \right)\right] \,,  \label{eq:loglambda2}
\end{align}
where $X_{I}=\mathcal{B}_I-\mathcal{E}_I$, $Y_{I}=\mathcal{C}_I+\mathcal{D}_I$, 
and $Z_{I}=\mathcal{B}_I+\mathcal{E}_I$.  

The calculation of the GLRT can be seen as a nested maximization problem, 
\begin{equation}\label{eq:maxie}
\text{GLRT}(\mathbf{r})=\max_{\lambda_i} \max_{\lambda_e} \Lambda(\mathbf{r};\lambda) \,.
\end{equation}
This split is meant to indicate that the whole model parameter set 
$\lambda$ can be divided into disjoint subsets classified as {\em extrinsic} 
(inner maximization) $\lambda_e=\lbrace\varphi_I\rbrace$, and {\em intrinsic} 
(outer maximization) $\lambda_i=\lbrace \alpha,\delta,\omega_{\text{gw}},\zeta,\iota,\psi,\varphi_0 \rbrace$. 
Usually, the separation is made such that the former can be maximized 
using analytical (or semi-analytical) methods, while the latter requires 
computationally expensive numerical optimization. It should be emphasized 
that the classification of parameters as extrinsic (computationally trivial) 
and intrinsic (computationally non-trivial) pertains to their role in the 
numerical procedure adopted for their estimation rather than their role 
in defining the the astrophysical signal.

\subsection{Maximization over extrinsic parameters}\label{sec:maxext}

The inner maximization of the GLRT over the extrinsic parameters 
(Eq.~\ref{eq:maxie}) leads to, 
\begin{equation}\label{eq:rooteq}
c^I_1\sin2\varphi_I + c^I_2\cos2\varphi_I + c^I_3\sin4\varphi_I + c^I_4\cos4\varphi_I=0  \,,
\end{equation}
where
\begin{subequations}\label{eq:coffe}
\begin{align}
c^I_1&=-\langle r^I|X_{I}\rangle_I + \langle X_{I}|Z_{I}\rangle_I \,, \\
c^I_2&=\langle r^I|Y_{I}\rangle_I - \langle Y_{I}|Z_{I}\rangle_I  \,,  \\
c^I_3&=\frac{1}{2}\big(\langle X_{I}|X_{I}\rangle_I - \langle Y_{I}|Y_{I}\rangle_I \big)  \,,  \\
c^I_4&=-\langle X_{I}|Y_{I}\rangle_I \,.
\end{align}
\end{subequations}

By defining $y=\cos2\varphi_I$, Eq.~\ref{eq:rooteq} can be transformed 
into a set of $N_p$ quartic equations 
\begin{equation}\label{eq:yeq}
ay^4+by^3+cy^2+dy+e=0
\end{equation}
where 
\begin{subequations}\label{eq:coff4quartic}
\begin{align}
a&=4(c_3^2+c_4^2)  \,, \\
b&=4(c_1c_3+c_2c_4) \,,  \\
c&=c_1^2+c_2^2-4(c_3^2+c_4^2) \,, \\
d&=-4c_1c_3-2c_2c_4  \,, \\
e&=c_4^2-c_1^2  \,.
\end{align}
\end{subequations}
Here, we have suppressed the pulsar index $I$ in Eq.~\ref{eq:yeq} and 
\ref{eq:coff4quartic} for clarity. 

A convenient numerical algorithm for solving quartic equations involves computing the 
eigenvalues of the $4\times4$ \textit{companion} matrix \citep{2002nrc..book.....P} 
\begin{equation}\label{Diagmatrix}
\mathbf{D}=\left( \begin{array}{cccc}
-\frac{b}{a} & -\frac{c}{a} & -\frac{d}{a} & -\frac{e}{a} \\
1 & 0 & 0 & 0 \\
0 & 1 & 0 & 0  \\
0 & 0 & 1 & 0 \\ 
\end{array} \right)\,.
\end{equation}
It is an upper Hessenberg matrix, for which the characteristic polynomial 
is Equation (\ref{eq:yeq}) with $y$ as the eigenvalue.  Hence, the set of 
its eigenvalues constitute the roots of the quartic equation. 

It is possible to get multiple real solutions (two or four) depending on 
the coefficients in Eq.~\ref{eq:coff4quartic}. Out of these solutions, we first 
select the ones whose absolute value is less than unity (since $y = \cos(2\varphi_I)$) 
and then select the one for which $\Lambda_I$ is greatest.  This ensures that the 
solutions for $\varphi_I$ found above also maximize the network log-likelihood 
ratio since it is just the sum over $\Lambda_I$. 

If no valid solution is found, then there is 
no turning point for $\Lambda_I$ in Eq.~\ref{eq:loglambda2}. For this case, 
the maximum of $\Lambda_I$ will appear at the boundary of the allowed 
region, i.e., $y=1$ ($\varphi_I=0$) and $y=-1$ ($\varphi_I=\pi/2$). 
We then evaluate $\Lambda_I$ at the boundary points and pick the one 
that gives the largest value.

\subsection{Maximization over intrinsic parameters} \label{sec:maxint}

The outer maximization of the GLRT over the intrinsic parameters (Eq.~\ref{eq:maxie}) 
requires a search for the global maximum over the remaining 7-D intrinsic parameter space  
$\lambda_i=\lbrace\alpha, \delta,\omega_{\text{gw}},\zeta,\iota, \psi, \varphi_0\rbrace$. 
This function is highly multi-modal due to the presence of noise in the data and 
degeneracies among the parameters. Deterministic local 
optimization fails to locate the global optimum in such a case and
a brute force grid search is computationally prohibitive for such a
large number of parameters. The only feasible approach
is to use algorithms that employ some type of a stochastic search scheme. 
As demonstrated in WMJ1, Particle Swarm Optimization (PSO)~\citep{eberhart1995new, 
2010PhRvD..81f3002W, 2012AstRv...7b..29M, 2012AstRv...7d...4M} 
provides a relatively straightforward approach to successfully addressing this problem.

PSO searches for the 
global optimum of a given fitness function using a stochastic
sampling scheme. The sample points, called ``particles", are iteratively
displaced according to the PSO dynamical equations. Relevant details 
of the PSO algorithm are provided in WMJ1. Since the present
optimization problem has a much lower dimensionality, one would expect that
the same PSO algorithm as used in WMJ1 would work here too. However, 
our initial tests showed that some tweaks were needed to achieve satisfactory 
performance. In order to describe these modifications, let us first recapitulate the PSO dynamical equations. 

Let $f(x)$ be a {\em fitness } function (i.e. the log likelihood ratio 
$\Lambda(\mathbf{r};\lambda$) in our case), where
$x\in S \subset \mathbb{R}^n$ and $S$ is called the {\em search space} 
and it is generally assumed to be a hypercube, $S = [a,b]\otimes[a,b]\otimes\ldots \otimes [a,b]$.
Let $x_i (k)$, $i = 1, 2,\ldots,N_{\rm part}$, be the
position of the $i^{\rm th}$ particle in a swarm of $N_{\rm part}$ particles at the iteration step $k$. The coordinates corresponding to $x_i(k)$ are $(x_{i,1}(k),\ldots,x_{i,n}(k))$.
Associated with each particle is the location, $p_i(k)$, called {\em pbest} (``particle best"), where the best fitness was found in its history.
\begin{eqnarray}
f\left(p_i(k)\right) & = & \max_{j = k,k-1,\ldots,0} f\left(x_i(j)\right)\;.
\label{updatepbest}
\end{eqnarray}
Associated with the swarm is  the location,  $g(k)$, called {\em gbest} (``global best"), 
where the best fitness was found by the swarm.
\begin{eqnarray}
f\left(g(k)\right) & = & \max_{j = 1,\ldots,N_{\rm part}} f\left(p_j(k)\right)\;.
\label{updategbest}
\end{eqnarray}
Given $x_i(k)$, $p_i(k)$ and $g(k)$, the following equations are used
to evolve the swarm.
\begin{eqnarray}
x_i(k+1) & = & x_i(k) +v_i(k);\label{positionupdate}\\
v_{i,j}(k+1) & = & \min\left(\max \left(y_{i,j}(k+1),-v_{\rm max}\right),v_{\rm max}\right)\;,\\
y_i(k+1) & = & w(k) v_i(k) + {\bf m}_{i,1}(p_i(k)-x_i(k))+
{\bf m}_{i,2}(g(k) - x_i(k))\;,\label{actualpsovelocity}
\end{eqnarray}
Randomness in the sampling is introduced through ${\bf m}_{i,p}$, $p = 1,2$, a diagonal matrix, ${\rm diag}(m_{p,i,1},\ldots,
m_{p,i,n})$, such that $m_{p,i,k}\sim U[0,c_p]$ is drawn from a
uniform distribution over $[0,c_p]$. The parameters $c_p$, $p = 1, 2$ and
the prescribed deterministic sequence $w(k)$ determine the 
extent to which continuing exploration of the search space is balanced by exploitation and focussing of the search around a good value at a given
iteration step. We set $w(k)$ to be a linearly decaying sequence
starting at 0.9 and ending at 0.4 at termination. At the termination of PSO, the highest fitness value found by the swarm, and the location of the particle with that fitness, make up the solution to the optimization problem. 

As in WMJ1, we use a modified form of the above iteration equations where
{\em gbest} is replaced by the best location, {\em lbest}, in a local 
neighborhood of each particle. We use the {\em ring topology} to determine
the neighborhoods: the particle indices are put on a circle and the neighborhood of each particle consists of $(m-1)/2$ particles
on each side with $m$ being the user specified size of each neighborhood.

The settings for the parameters of the PSO algorithm outlined above
 are retained from WMJ1: $N_{\rm part} = 40$, $c_1 = c_2=2.0$,  $m = 3$, $v_{\rm max} = (b-a)/5$, $v_{\rm max}^\prime = (b-a)/2$, $w(k) = 0.9 - 0.5(k/(N_{\rm iter}-1))$, where $N_{\rm iter}=2000$ is the total number of iterations. In addition to fixing the PSO parameters, the behavior of particles crossing the boundary of $S$ is handled using the ``let them fly" boundary condition in which the fitness of the particle is simply set to $-\infty$ while it is outside $S$. The main modification to the PSO algorithm in this paper 
 is the introduction of a local optimization of the {\em gbest} position, using
 the Nelder-Mead algorithm \citep{2002nrc..book.....P}, that is performed only when {\em gbest} changes. We
 believe that the maximization over the pulsar phases leaves behind a fitness function
 that has ridge-like features in it. 
 This expectation is based on similar behavior of the 
 fitness function, after initial phase maximization, in the case of compact binary inspiral signals for ground-based
 searches. The use of local optimization then moves the {\em gbest} location
 along these long ridges to better values more efficiently than a pure random move. 
 However, a systematic study of these ideas is postponed to a future work.

To increase the probability of successfully converging to within a 
sufficiently small neighborhood of the global maximum,  multiple independent 
runs of PSO are made on the same data segment. Being
mutually independent, these runs can be executed using simple parallelization on 
a multi-processor machine.
Unlike the case of WMJ1, where the computational cost of evaluating
the fitness function was relatively higher and only one run of PSO per data realization was feasible,  
we are able to execute 8 independent runs for each
data realization in the present case. 

For simulated data, it is possible to gauge successful convergence to the global maximum by 
comparing the best fitness found with its value at the true signal location: 
the former should always be higher than the latter. This test is passed by PSO 
in all the cases discussed in the next section. 

\section{Applications} \label{sec:applications}

We illustrate the above algorithm (hereafter referred to as MaxPhase) using 
simulated data corresponding to a PTA configuration 
adopted in \citet{2014PhRvD..90j4028T} (see the paper and the references therein 
for ephemerides of the nine pulsars in the network). In all of the cases considered below, 
the source is a SMBHB in a circular orbit which is located at Right 
Ascension $\alpha=1.0$ rad ($3~\text{hr}~49~\text{min}$) and declination 
$\delta=0.5$ rad ($28^{\circ}.7$).  The orbital angular 
angular frequency $\omega=1.96~\text{rad yr}^{-1}$ ($\omega_{\text{gw}}=3.93~\text{rad yr}^{-1}$), 
the initial phase $\varphi_0=2.89~\text{rad}$ ($165^{\circ}.6$), the inclination angle 
$\iota=0.5~\text{rad}$ ($28^{\circ}.6$) and the polarization angle 
$\psi=0.5~\text{rad}$ ($28^{\circ}.6$) 
are also set to be the same values as in \citet{2014PhRvD..90j4028T}. 
The span of the simulated observation is 14.9 years, with uniform 
biweekly cadence leading to the same number of samples 
$N_I=389$ for each pulsar. The signal induced by this GW 
source is calculated for each pulsar in the PTA following Eq.~\ref{eq:residualphi}. 
Independent realizations of white Gaussian noise are added to the signal 
for each pulsar, with the noise standard deviation $\sigma^I$  for a given 
pulsar set equal to its timing residual rms (we used the same level of noise 
as in WMJ1).  To characterize the strength 
of the signal in the data, we use the network SNR of the signal defined as 
\begin{equation}\label{eq:snr}
\rho_n=\left(\sum_{I=1}^{N_p}\langle s^I|s^I \rangle _I\right)^{1/2}
=\left(\sum_{I=1}^{N_p}\sum_{k=1}^{N_I}\left(\frac{s^I_k}{\sigma^I}\right)^2\right)^{1/2} \,.
\end{equation}
We choose $\rho_n=100$, 30, 8 to represent 
the strong, moderate, and weak signal scenarios respectively. 
For each scenario, 200 independent realizations of data are generated. 
The results from each scenario are discussed in the following sub-sections. 
Although not required from the point of view of signal analysis, 
these S/N choices can be associated with astrophysical parameters 
for concreteness. For example, the S/N values above in descending 
order could arise from a SMBHB system that has a chirp mass 
$\mathcal{M}_c\approx 10^9~M_{\odot}$, an orbital period of $P=3.2$ yrs, 
and that is located at a distance approximately 10, 30 and 125 Mpc 
from Earth, respectively.  
As already mentioned in Sec.~\ref{sec:datamodel}, we ignore the evolution 
of the binary orbital frequency, which is a reasonable assumption for the purpose 
of studying the performance of the algorrithm, although this assumption 
can become invalid in the late stage of the SMBHB evolution. 

Fig.~\ref{fig:PSOvsTrue} compares the log likelihood ratio found by the 
MaxPhase algorithm with its value for the true signal parameters. 
For each of the three network S/N, we can see that the former is greater 
than the latter for all realizations. This is the least one expects from 
any viable estimation algorithm and we see that the 
MaxPhase algorithm passes this basic test. 

\begin{figure}[H]
\begin{center}
\begin{tabular}{cc}
\epsfig{file=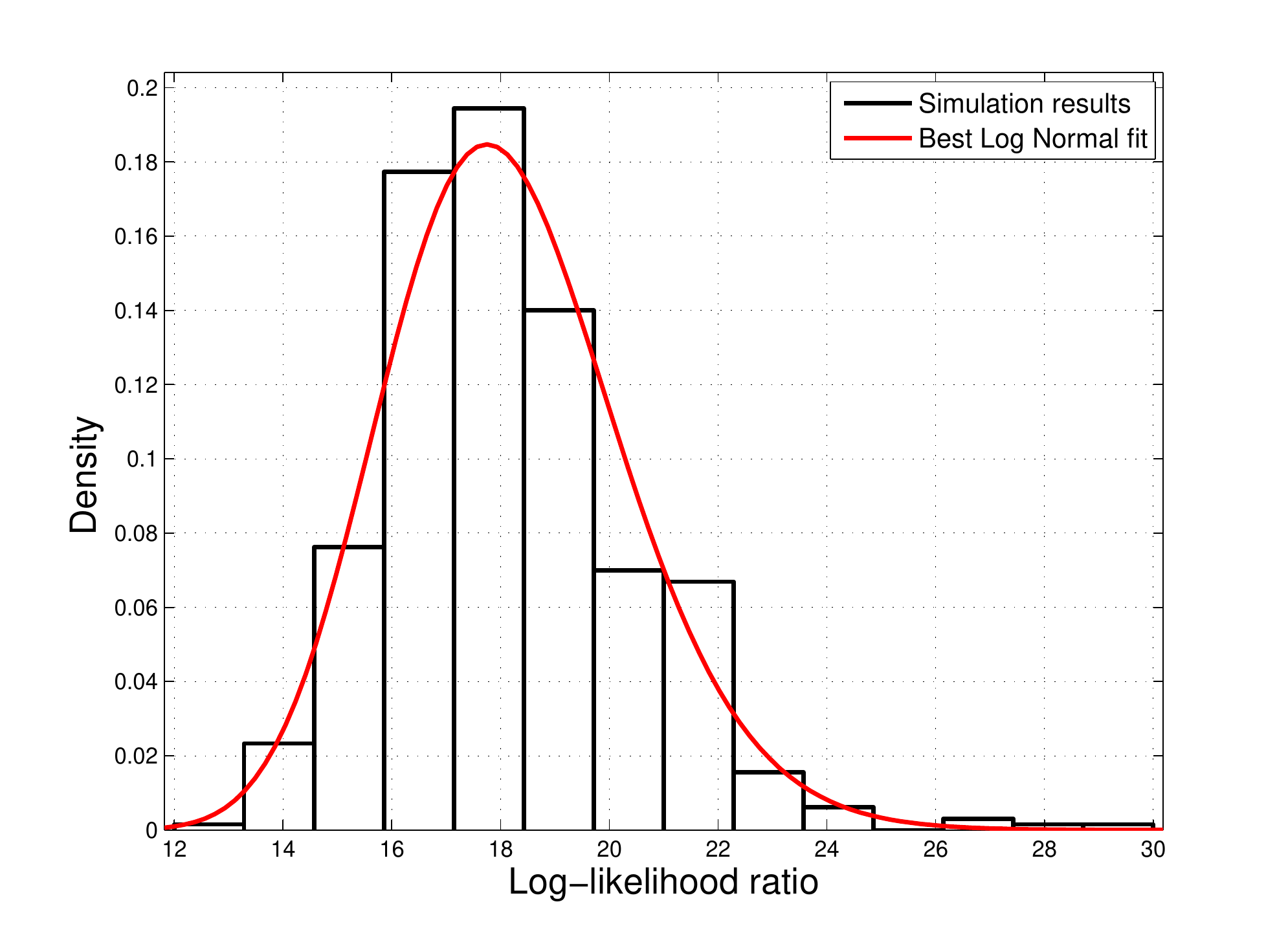,width=0.45\linewidth,clip=} &
\epsfig{file=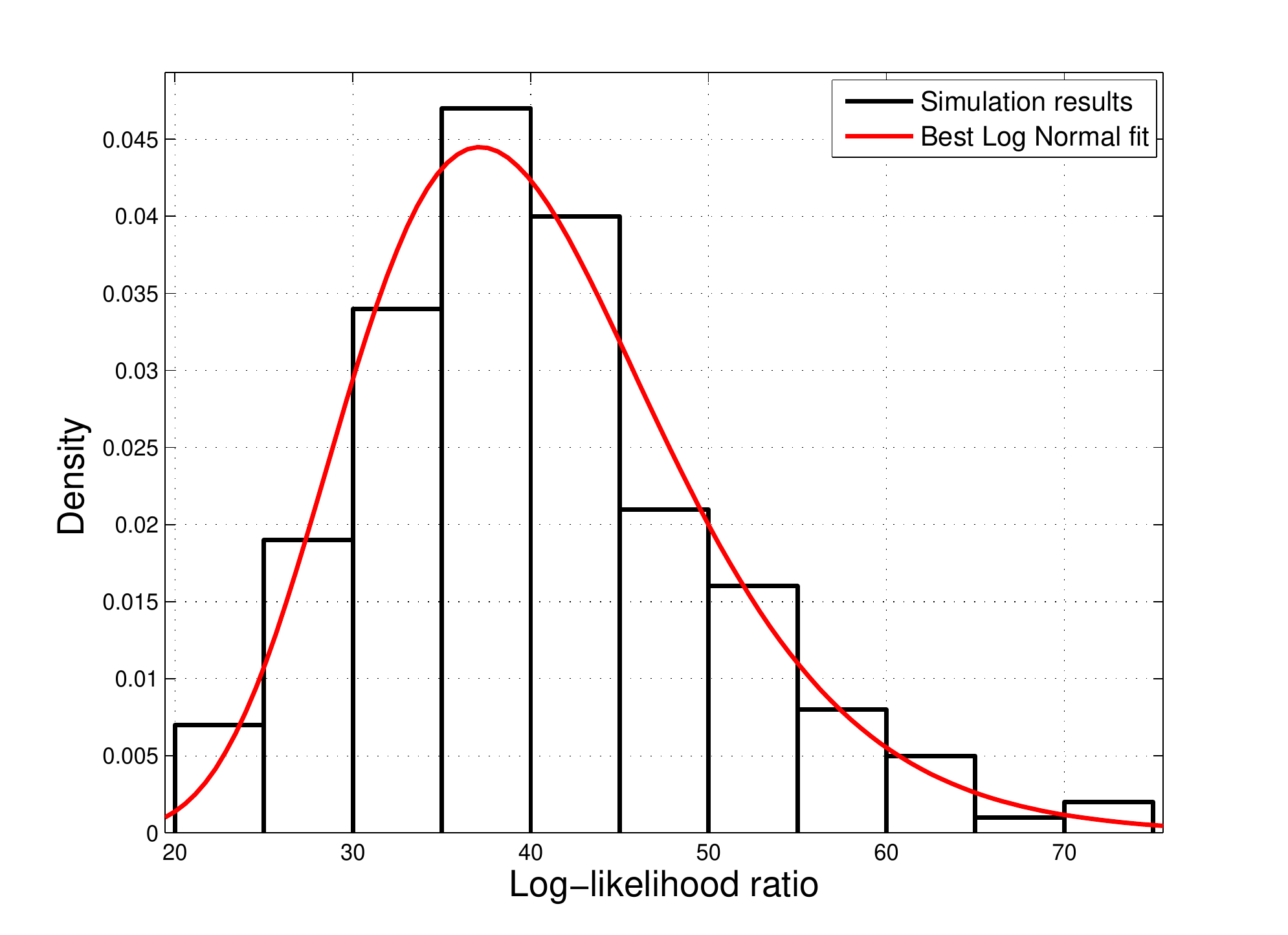,width=0.45\linewidth,clip=} \\
\epsfig{file=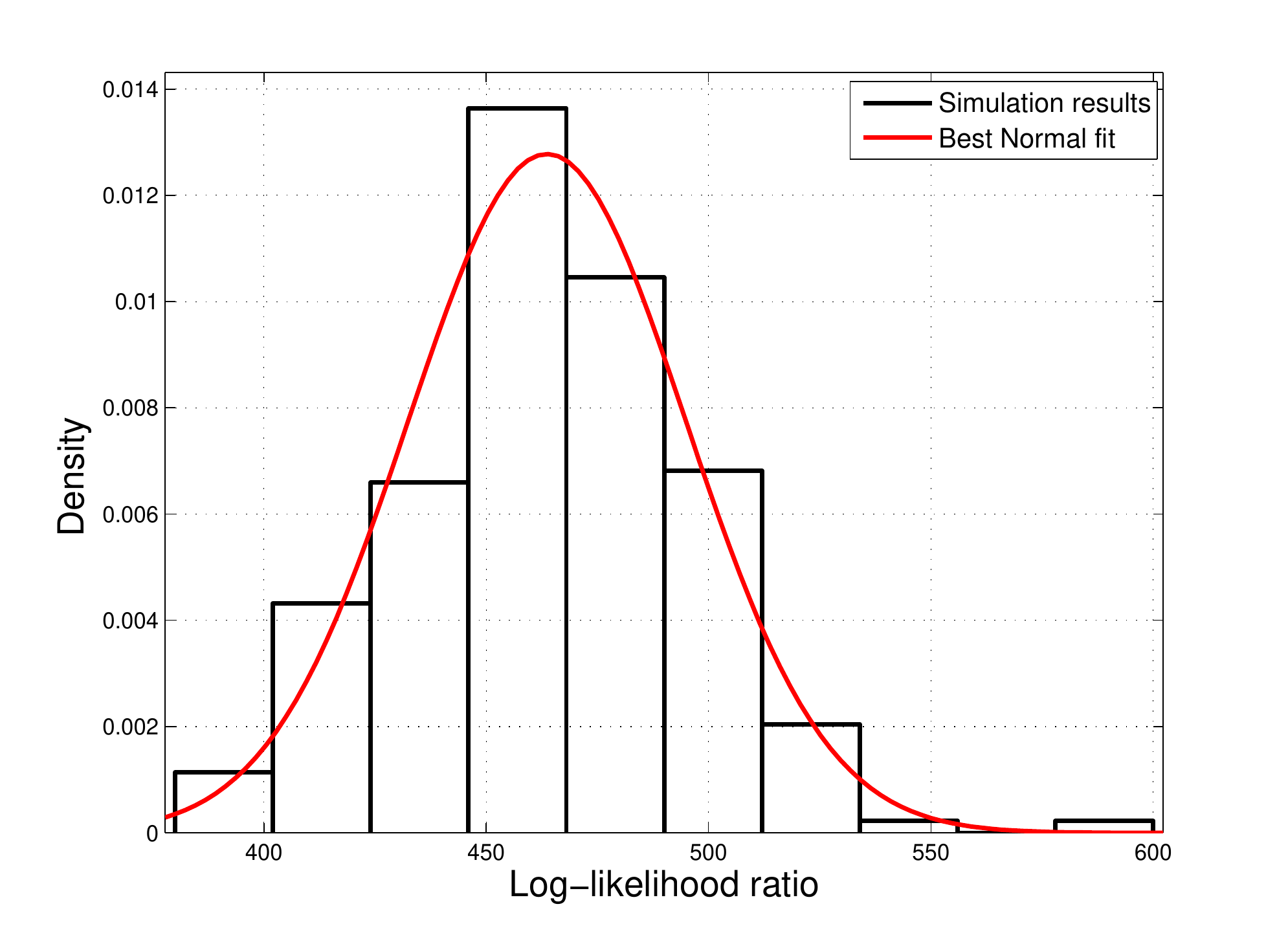,width=0.45\linewidth,clip=} &
\epsfig{file=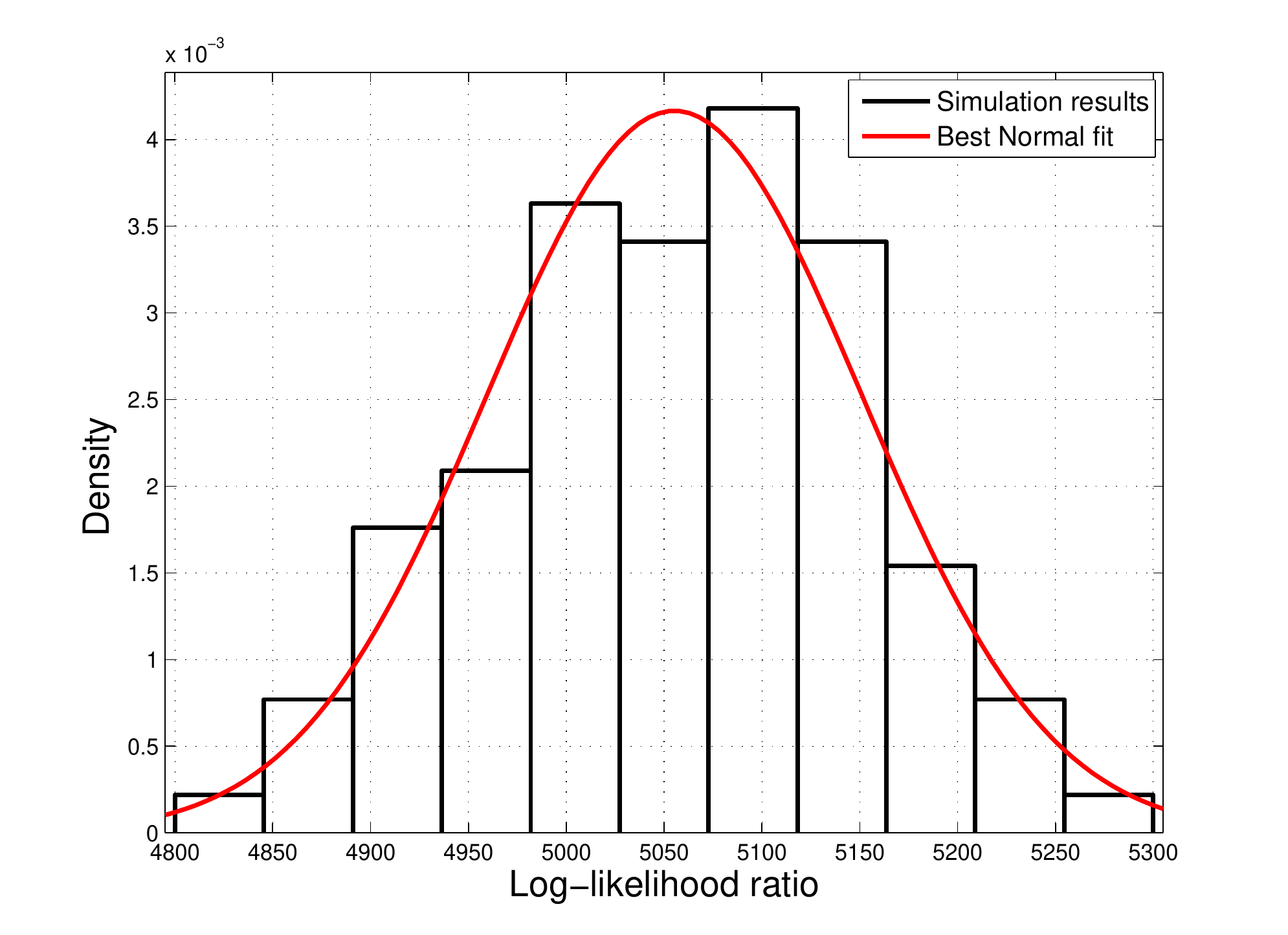,width=0.45\linewidth,clip=}
\end{tabular}
\end{center}
\caption{Histograms of the detection statistic normalized by the total 
number of trials. The histogram in the upper left panel 
is for the $\mathcal{H}_0$ case; the histogram in the upper right panel is for $\rho_n=8$ case; 
the histogram in lower left panel is for the $\rho_n=30$ case; the histogram in the lower 
right panel is for the $\rho_n=100$ case. The red curve in the each panel shows the 
best fit distribution. These are $\ln\mathcal{N}(\mu=2.89,\sigma=0.12)$, $\ln\mathcal{N}(\mu=3.67,\sigma=0.23)$, 
$\mathcal{N}(\mu=463.7,\sigma=31.2)$, and $\mathcal{N}(\mu=5055.3,\sigma=95.8)$ respectively.}
\label{fig:hist0123}
\end{figure}

\begin{figure}[ht]
\centerline{\includegraphics[width=1.2\textwidth]{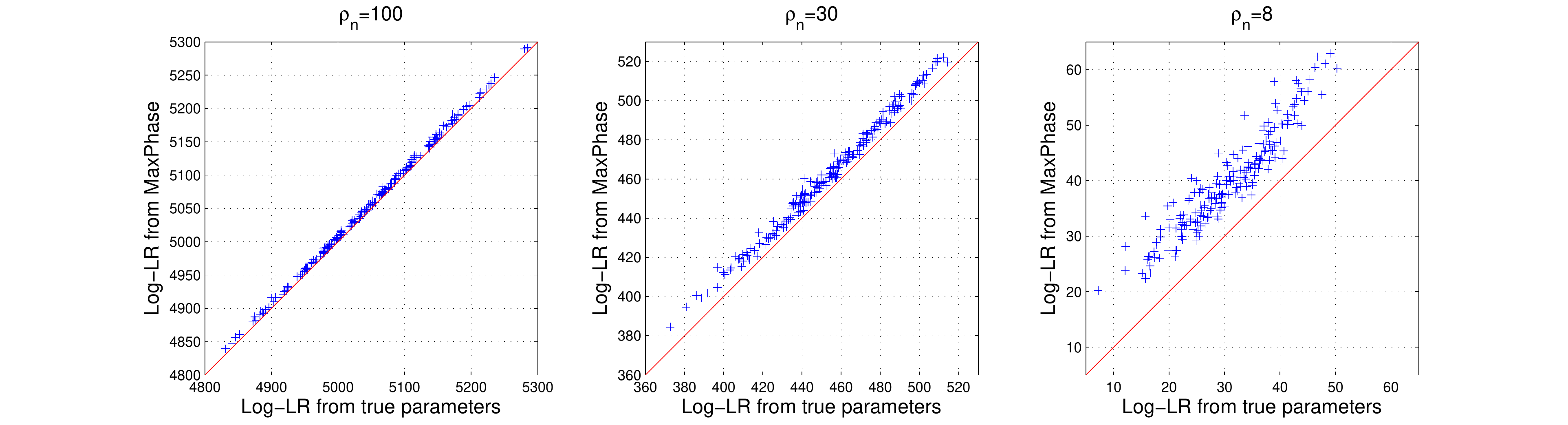}}
\caption{Log likelihood ratio values obtained from MaxPhase v.s. Log likelihood 
ratio for the true signal. From left to right, the panels correspond to the network 
S/N $\rho_n=100$, 30, 8 scenarios, respectively. There are 200 data realizations 
for each scenario.}
\label{fig:PSOvsTrue}
\end{figure}

To obtain the threshold for detection or to set upper limits, 
the distribution of the detection statistic under $\mathcal{H}_0$ is required. 
This involves finding the distribution of the maximum of the log 
likelihood ratio $\Lambda$.  We use Monte-Carlo simulation with 
500 independent noise-only realizations of data to directly estimate 
this distribution. Figure \ref{fig:hist0123} shows 
the distributions of the detection statistic GLRT$(\mathbf{r})$ under 
the noise-only case and under the three different signal scenarios. 
The histograms for the $\mathcal{H}_0$ and $\rho_n=8$ cases can be fitted well 
by the Log-Normal distribution $\ln\mathcal{N}(\mu,\sigma)$. The 
distribution converges to a normal distribution $\mathcal{N}(\mu,\sigma)$ 
as the signal strength increases.

\subsection{Strong signal} \label{sec:strong}

\begin{figure}[H]
\centerline{\includegraphics[width=1.2\textwidth]{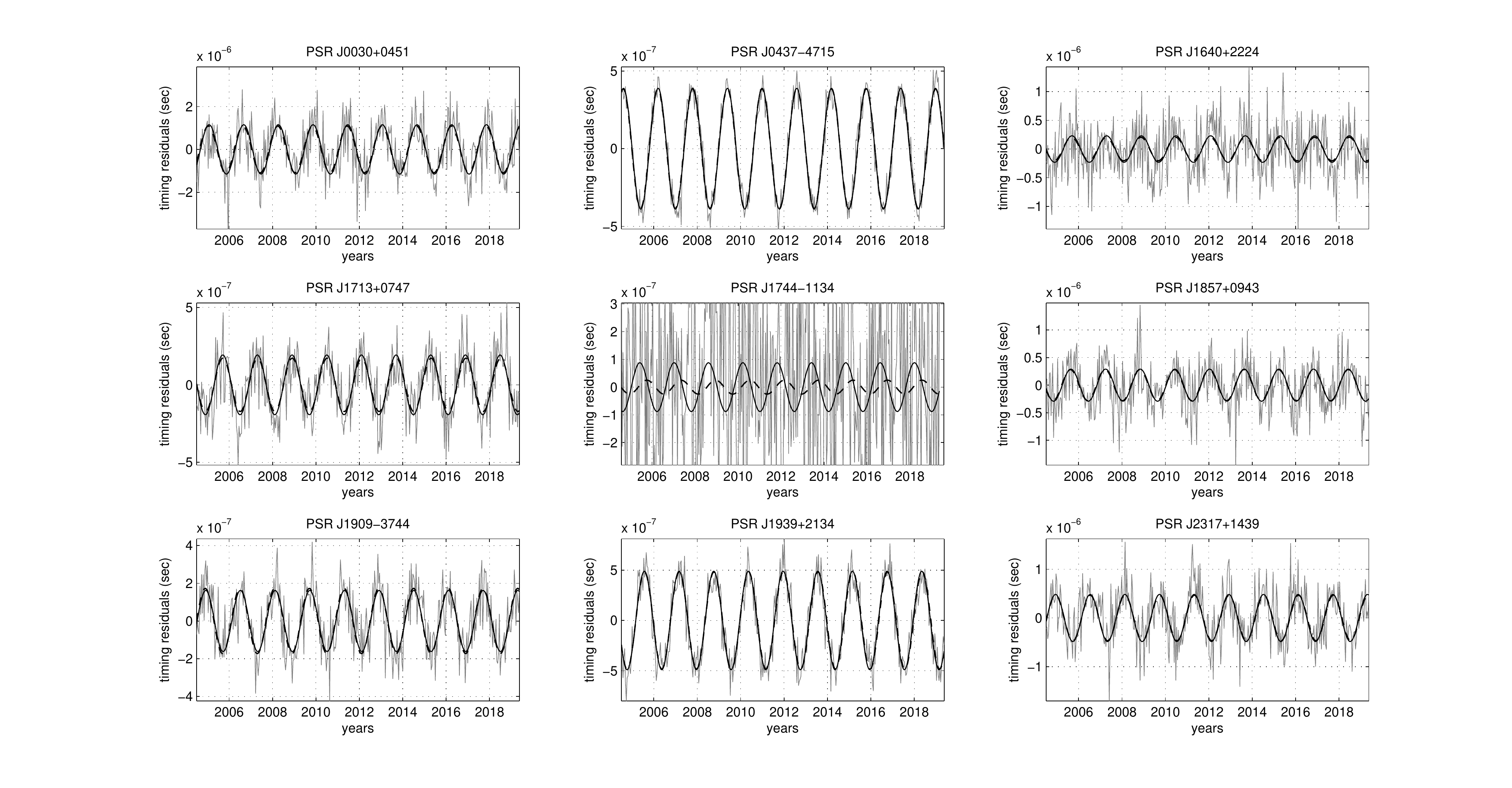}}
\caption{Data realization showing the simulated timing residuals (thin gray line) 
and signal (dash black line) for all pulsars. The network S/N $\rho_n=100$. 
The reconstructed signals are shown as solid curves. For most pulsars, except J1744--1134, the 
true and reconstructed signal are almost indistinguishable from each other. 
For PSR J1744--1134, we have zoomed into the noise so that the signal can be seen clearly.}
\label{fig:snr3_rlz112}
\end{figure}

In this case, the network S/N $\rho_{n}=100$.
Figure~\ref{fig:snr3_rlz112} shows a typical realization of the simulated 
timing residuals for the nine pulsars (thin gray line). The magnitude and the 
phase of the noise-free timing residual (black dashed line) depend 
on the location and distance of the source and pulsar in 
the array. In this strong signal scenario, the signal in most of the pulsars 
is comparable to or even stronger than the respective noise. 
The reconstructed signal is 
obtained by Equation~\ref{eq:residual2} or \ref{eq:residualphi} in which 
the input extrinsic and intrinsic parameters are the ones estimated by MaxPhase.

As seen from Figure~\ref{fig:snr3_rlz112}, the estimated signal is 
indistinguishable from the injected one for all the pulsars except PSR 
J1744--1134 (separation angle is $150^{\circ}$) which contains the 
weakest signal and contributes insignificantly to the detection statistic. 
Figure~\ref{fig:hist0123} shows the distribution of the detection 
statistic values under the null ($\mathcal{H}_0$) and alternative 
($\mathcal{H}_\lambda$) hypotheses. Comparing the distributions for 
null and $\rho_n=100$, it is clear that the detection 
probability $Q_d$ is nearly unity if the threshold for claiming a detection 
is chosen as the highest value for the null case. Since we have 
used 500 realizations for $\mathcal{H}_0$, the false alarm probability for this 
choice is approximately $2\times10^{-3}$. 

\begin{figure}[H]
\centerline{\includegraphics[width=1.0\textwidth]{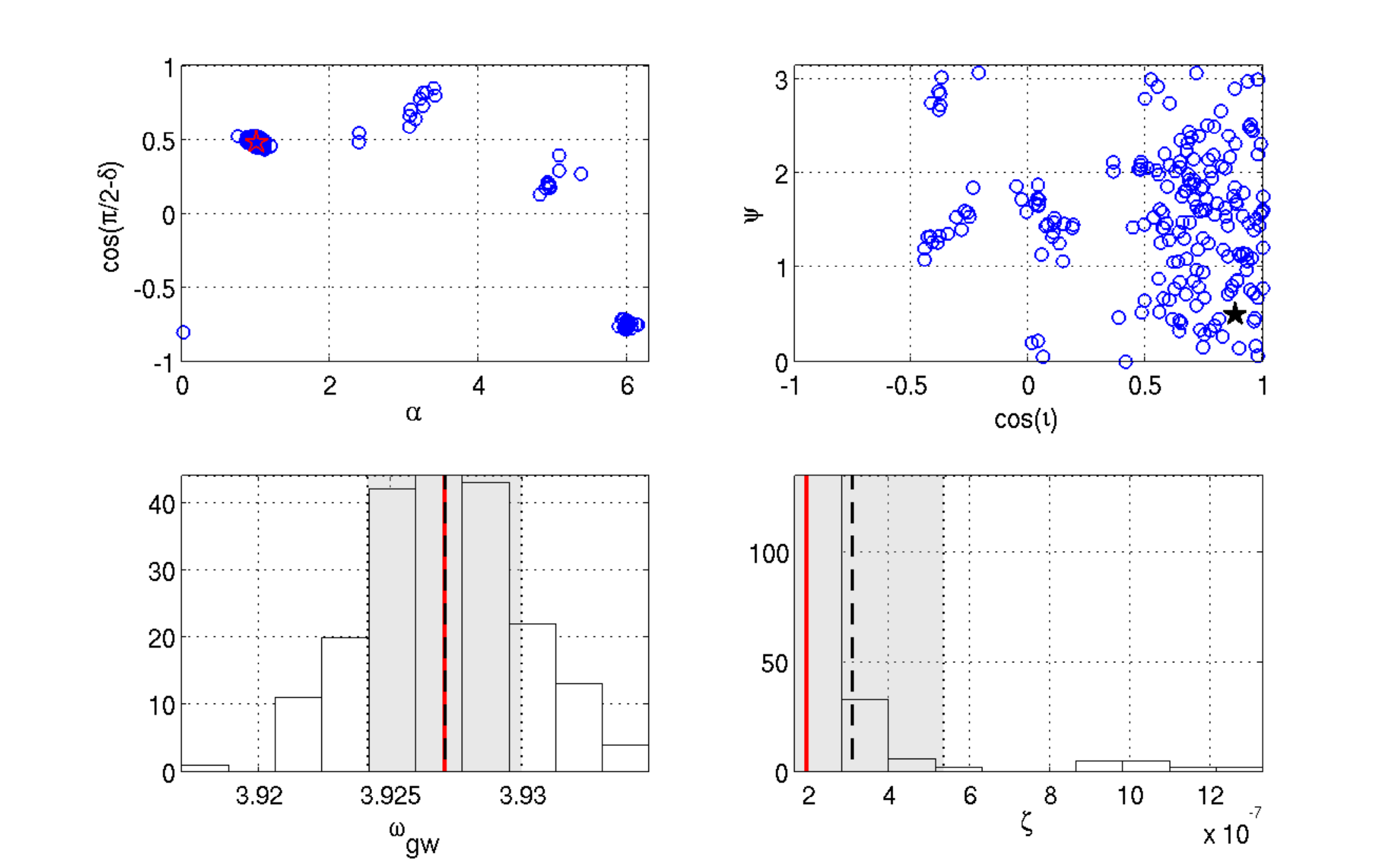}}
\caption{Two-dimensional scatter plots (top) and histograms (bottom) of estimated 
parameters for network S/N $\rho_n=100$. The star and the red vertical line 
mark the true values of the parameters. 
The dashed vertical line marks the mean value, and the shaded area covers 
the one-sigma region around the mean. The total number of trials is 200.}
\label{fig:hist_snr3}
\end{figure}

Figure~\ref{fig:hist_snr3} shows the distributions of estimated parameters 
$\lbrace\alpha,\delta,\zeta,\iota,\psi,\omega_{\text{gw}}\rbrace$ that are 
astrophysically interesting. The distributions were estimated 
from 200 independent data realizations. 
For the sky localization, most of the estimated locations are very close to 
the true one. However, for 43 out of the 200 realizations, 
the estimated locations appear to fall on some secondary maxima located along 
an arc. Similarly, the scatter for $\iota$ and $\psi$ is larger than expected. 
In contrast, the true values of $\omega_{\text{gw}}$ and $\zeta$ are well within the 
one-sigma uncertainty of $2.9\times 10^{-3}~\text{rad}\cdot\text{yr}^{-1}$ 
and $6.11\times 10^{-7}~\text{sec}$, respectively, calculated from the 200 realizations.

\subsection{Moderate signal} \label{sec:moderate}

\begin{figure}[H]
\centerline{\includegraphics[width=1.2\textwidth]{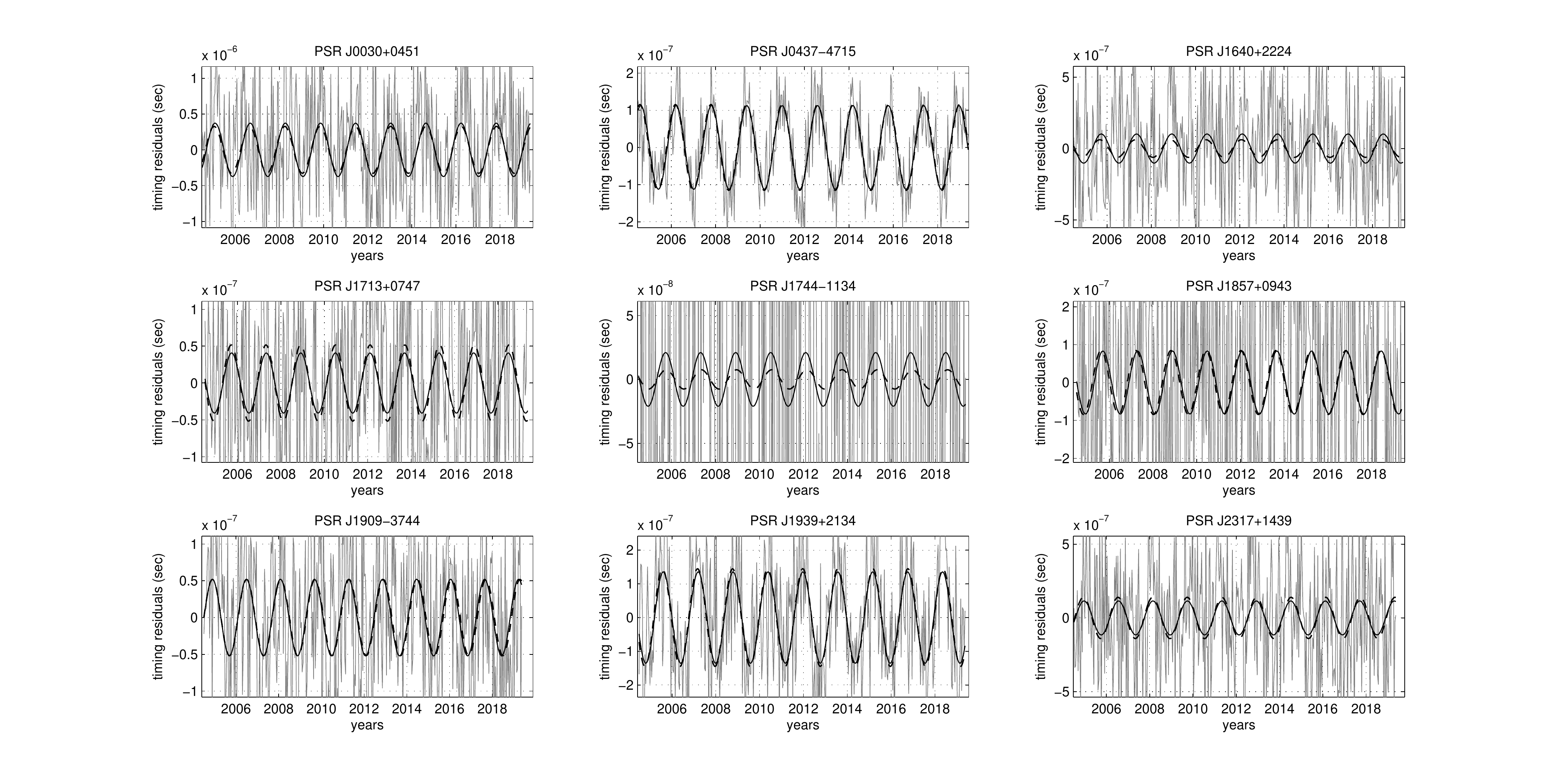}}
\caption{Data realization showing the simulated timing residuals (thin gray line) 
and signal (dash black line) for all pulsars. The network S/N is $\rho_n=30$. 
The reconstructed signals are shown as solid curves. For some pulsars, such as 
PSR J1744--1134 and J1857+0943, we have zoomed into the noise in the subplots, so 
that the signal can be seen clearly.}
\label{fig:snr2_rlz63}
\end{figure}

\begin{figure}[H]
\centerline{\includegraphics[width=1.0\textwidth]{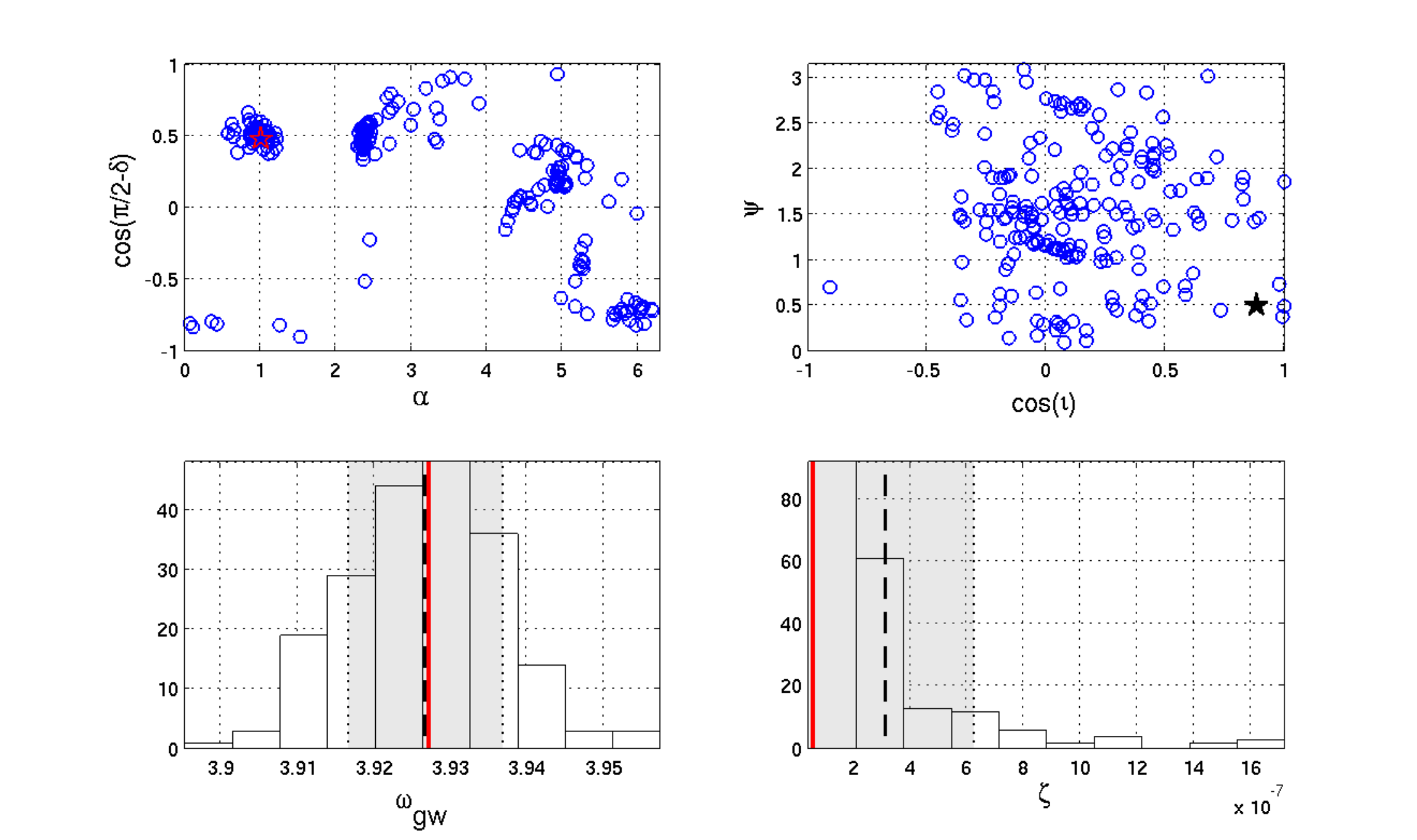}}
\caption{Two-dimensional scatter plots (top) and histograms (bottom) of estimated 
parameters for network S/N $\rho_n=30$. The star and the red vertical line 
mark the true values of the parameters. The dashed vertical line marks the 
mean value, and the shaded area covers the one-sigma region around the mean. 
The total number of trials is 200.}
\label{fig:hist_snr2}
\end{figure}

Figure~\ref{fig:snr2_rlz63} shows a realization of the simulated data 
for a network S/N $\rho_{n}=30$. 
The noise is now seen to be stronger than the signal in most of the pulsars. 
The recovered signal continues to agree with the injected one quite well. Note that for 
PSR J1744--1134, J1713+0747 and J1640+2224, the deviation from the true 
signals is mainly in the amplitude, while the offset in phase is not 
significant. From the distribution of the detection statistic in Fig.~\ref{fig:hist0123}, 
the detection probability is still practically unity for a detection 
threshold with an approximate false alarm probability of $2\times 10^{-3}$.  
In Figure~\ref{fig:hist_snr2}, 
we see more clearly that the sky locations are centered on the same 
secondary maxima as in the $\rho_n = 100$ case (Fig~\ref{fig:hist_snr3}) 
but with an increased scatter around each. We also note that the bias in 
the estimation of the inclination and polarization angle is increased. 
The one-sigma uncertainties for $\omega_{\text{gw}}$ and $\zeta$ increase 
to $0.01~\text{rad}\cdot\text{yr}^{-1}$ and $8.63\times 10^{-7}~\text{sec}$, 
respectively. The increase in the errors is roughly consistent with their 
expected linear dependence on network S/N.

\subsection{Weak signal} \label{sec:weak}

\begin{figure}[H]
\centerline{\includegraphics[width=1.2\textwidth]{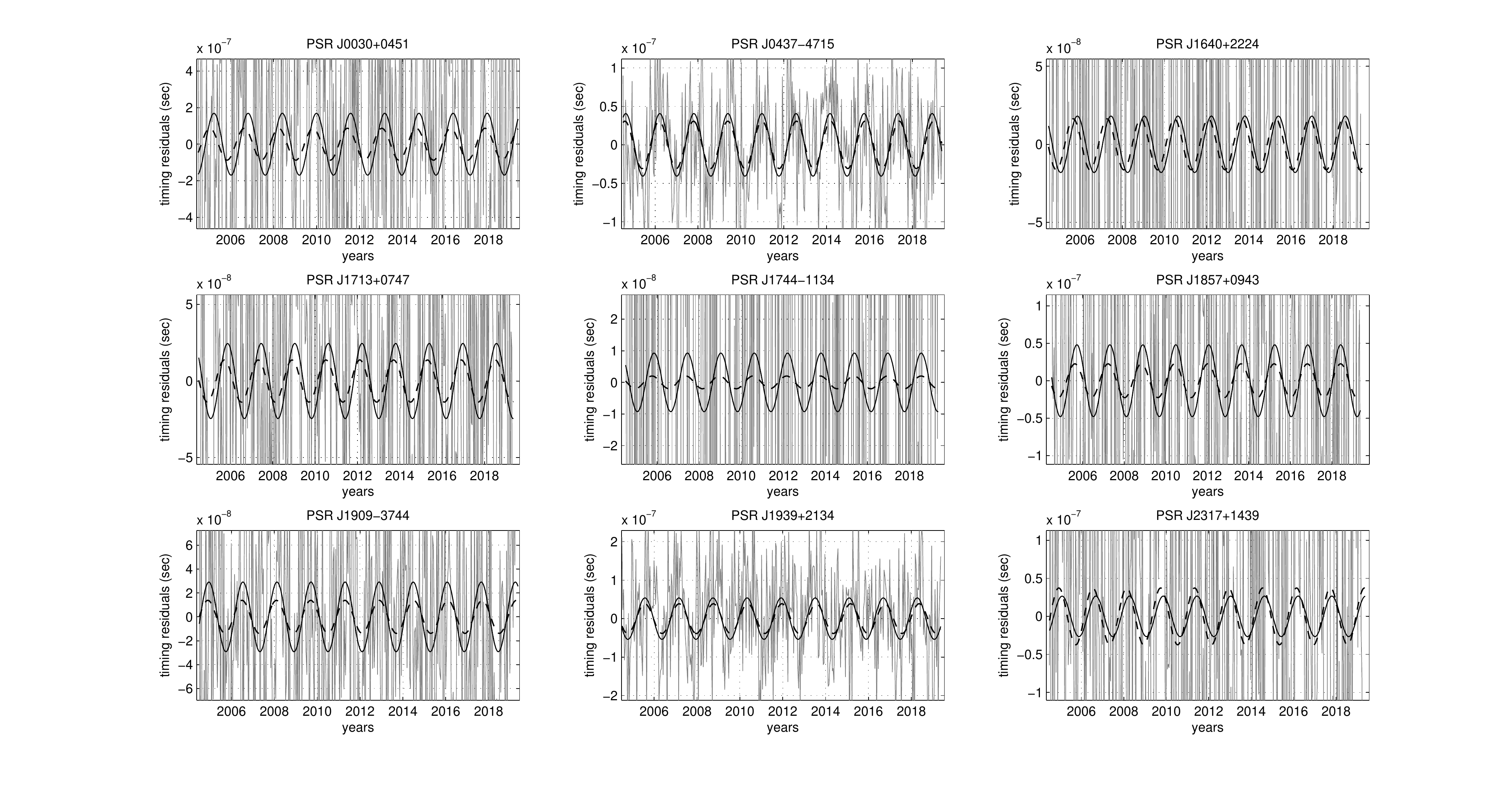}}
\caption{Data realization showing the simulated timing residuals (thin gray line) 
and signal (dash black line) for all pulsars. The network S/N is $\rho_n=8$. 
The reconstructed signals are shown as solid curves. For most pulsars, we have 
zoomed into the noise in the subplots, so that the signal can be manifested.}
\label{fig:snr1_rlz174}
\end{figure}

\begin{figure}[H]
\centerline{\includegraphics[width=1.0\textwidth]{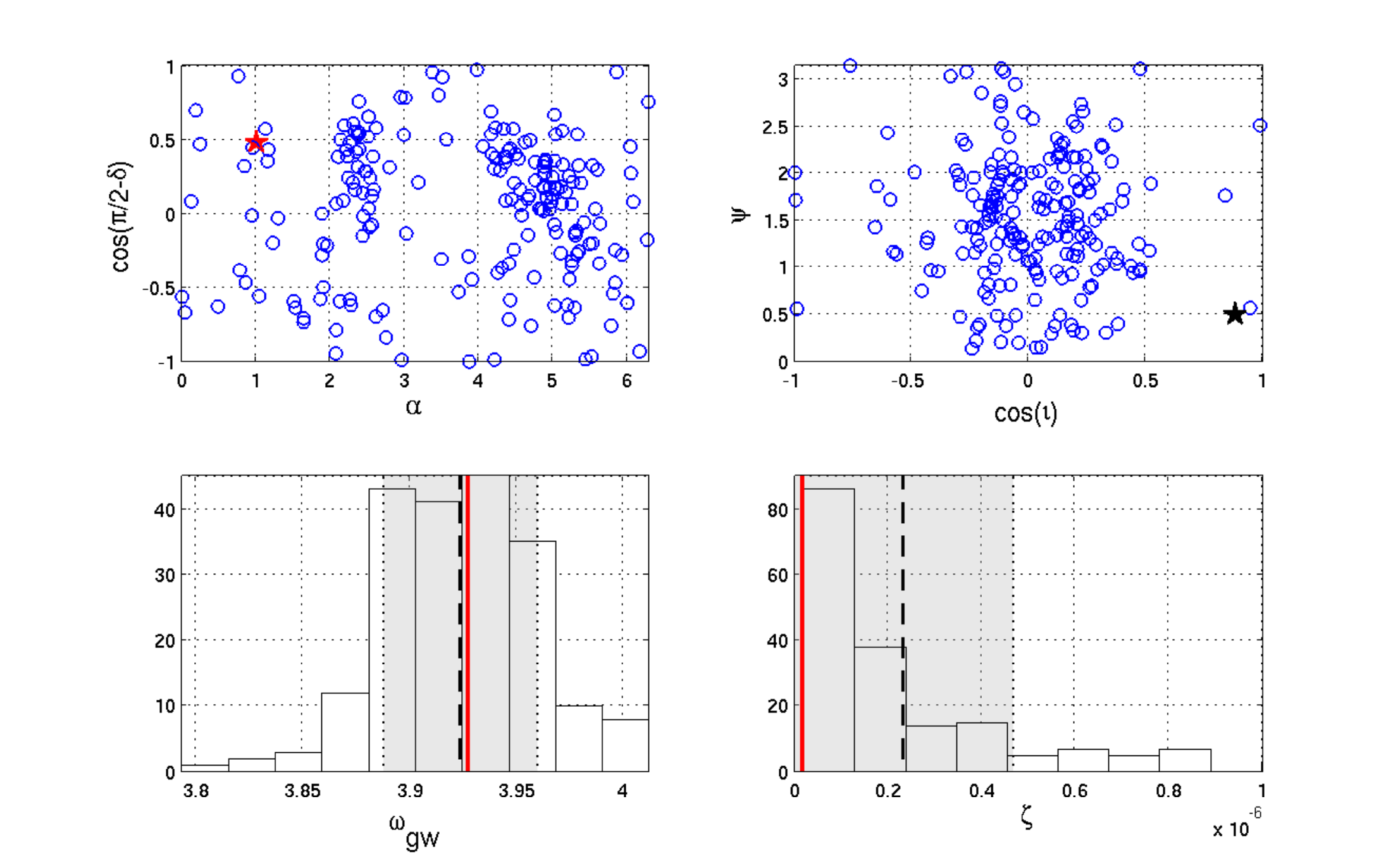}}
\caption{Two-dimensional scatter plots (top) and histograms (bottom) of estimated 
parameters for network S/N $\rho_n=8$. The star and the red vertical line 
mark the true values of the parameters. The dashed vertical line marks the 
mean value, and the shaded area covers the one-sigma region around the mean. 
The total number of trials is 200.}
\label{fig:hist_snr1}
\end{figure}

In this case, the network S/N $\rho_{n}=8$ corresponds to a weak and barely detectable signal. 
This is also the network S/N used in \citet{2014PhRvD..90j4028T}. 
Figure~\ref{fig:snr1_rlz174} shows one of the realizations of the simulated timing residuals. 
In this scenario, the noise dominates the signal in all pulsars. This illustrates 
the most likely situation with the current level of timing precision obtained in 
pulsar timing arrays. Even though the noise is loud, the recovered signals have deviations 
mainly in the amplitude (usually biased towards a larger value), while the offset in the phase is tolerable. 
In Figure~\ref{fig:hist_snr1}, the scatter of the sky location becomes larger, 
but the presence of secondary maxima seen in the previous cases is still discernible. 
However, now the true location attracts the least number of trial values. 
The bias in the estimation of the inclination and polarization angle is now much 
clearer. The uncertainties in $\omega_{\text{gw}}$ and $\zeta$ are 
$0.036~\text{rad}\cdot\text{yr}^{-1}$ and $2.11\times 10^{-6}~\text{sec}$, again 
roughly consistent with the expected linear dependence on network S/N. 
From Figure~\ref{fig:hist0123}, the detection probability is $Q_d\simeq 0.86$ 
if we choose the detection threshold to be the largest value of 
the noise-only distribution. In this case, the signal is still large 
enough to be detected, although it cannot be localized at all.

\subsection{Comparison with other algorithms} \label{sec:comp}

\begin{figure}[H]
\centerline{\includegraphics[width=1.3\textwidth, angle=0]{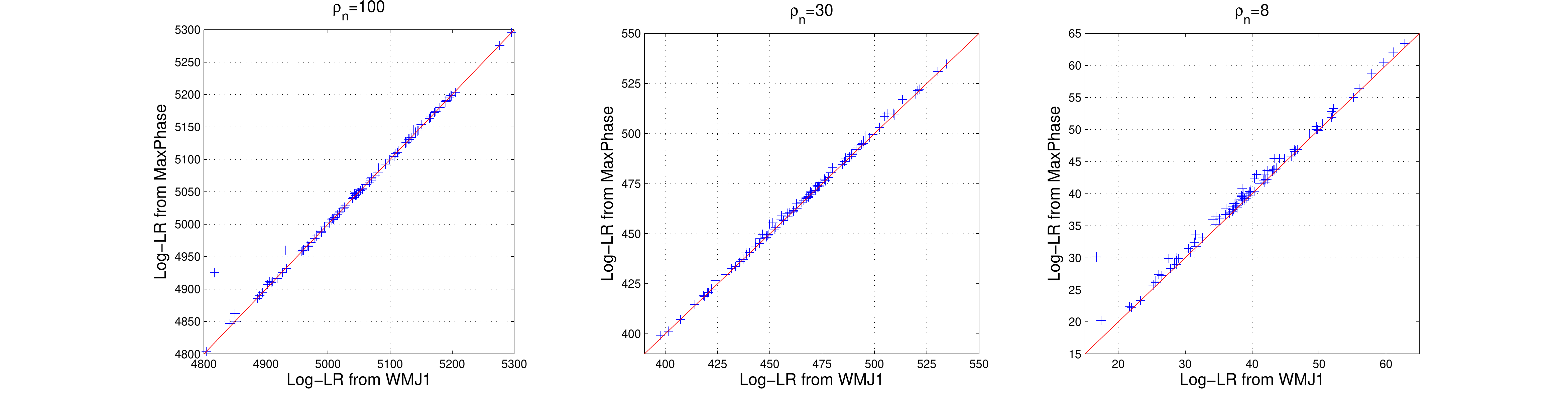}}
\caption{In each panel, Log likelihood ratio values from MaxPhase algorithm v.s. WMJ1 algorithm are shown 
for three scenarios with $\rho_n=100$, $30$, and $8$, respectively. The number of independent 
data realization is 100. In almost all trials, the log-likelihood ratios are seen to be higher for the 
MaxPhase algorithm.}
\label{fig:A1A2fit}
\end{figure}

In Figure~\ref{fig:A1A2fit}, we show the log likelihood ratio from MaxPhase versus 
those from WMJ1 for a subset of 100 data realizations chosen randomly from the set 
used for the simulations reported above.  We can see that for most realizations 
in each of the three signal strength scenarios, the former can find a marginally 
larger (better) log likelihood ratio than the latter, which suggests that MaxPhase 
can achieve a greater detection probability than WMJ1 for a given detection threshold. 

\begin{figure}[H]
\centerline{\includegraphics[width=1.3\textwidth, angle=0]{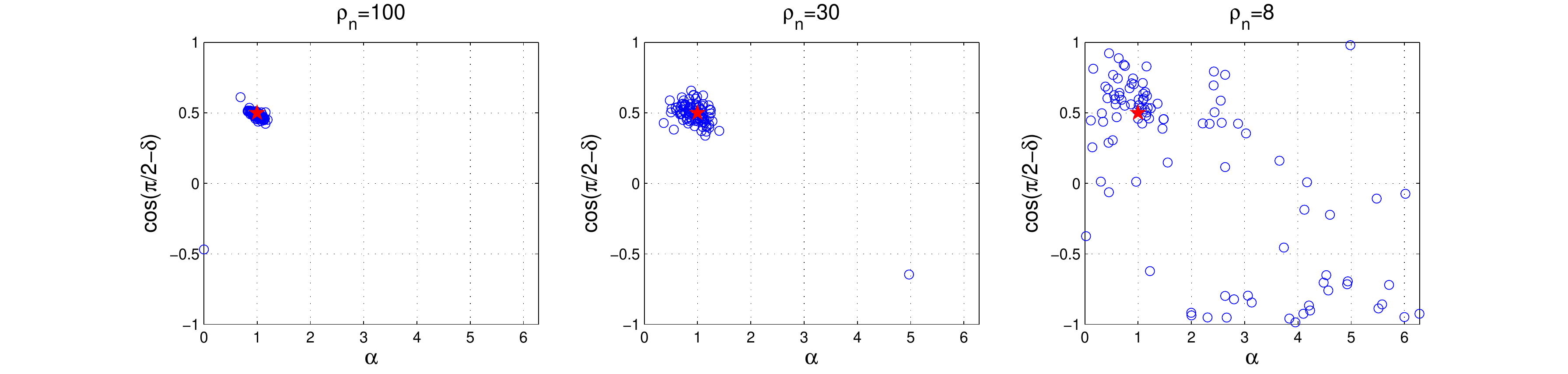}}
\caption{In each panel, blue circles show the estimated sky locations for the source, which are 
obtained from the WMJ1 algorithm for a PTA consisted of 9 pulsars. A red star marks the true 
location of the source used in the simulation. The x-axis represents Right Ascension and the 
y-axis represents the declination. The total number of independent data realization is 100. 
The panel on the right may be compared with Fig. 5 of \citet{2014PhRvD..90j4028T}. }
\label{fig:A1scat100}
\end{figure}

Comparing parameter estimation performance, Figure~\ref{fig:A1scat100} gives 
the estimated sky locations from WMJ1. For the $\rho_n=100$ case, the sky localization 
is very similar to the corresponding one in Figure~\ref{fig:hist_snr3}, except 
that there are no secondary maxima. With the decreasing of 
$\rho_n$ to 30 and 8, the sky localization scatter increases but it still 
appears uni-modal and concentrated around the true value.

Comparing our results for MaxPhase and WMJ1 with those of the Bayesian 
method \citep{2014PhRvD..90j4028T}, we make the following observations. 
From the receiver operating characteristics (ROC) curve reported in Fig. 6 
of \citet{2014PhRvD..90j4028T}, the detection probability appears to be close 
to unity for $\rho_n=8$ case at the lowest false alarm probability of 0.01 
used in that paper. The corresponding 
detection probability from MaxPhase is $97.5\%$ (and rapidly approaches unity for 
higher FAP). Thus, the detection performance of MaxPhase is comparable 
to that of the Bayesian method. The distribution of the estimated parameters 
in the Frequentist case can be compared more reliably with the distribution of the 
\textit{maximum-a-posteriori} value of the parameters in the Bayesian method. We 
have picked the same source parameters as in \citet{2014PhRvD..90j4028T}, 
so the comparison is straightforward. Although there are differences between 
the two analyses, such as the use of irregularly versus regularly sampled 
data, they should not impact the comparison too much. 
From Figure~\ref{fig:hist_snr1} (MaxPhase), Figure~\ref{fig:A1scat100} (WMJ1) 
in this paper and Figure~5 (Bayesian) in \citet{2014PhRvD..90j4028T}, 
we see that for $\rho_n=8$ case (the only case considered in \citet{2014PhRvD..90j4028T}), 
the estimated sky location by MaxPhase is inferior to the Bayesian method, 
while the results from WMJ1 and the Bayesian method are qualitatively comparable. 

Regarding computational costs, MaxPhase takes 6.7 min on average to complete 
one PSO run for each data realization on a single processor core, while the WMJ1 
algorithm takes 89 min. 
As far as obtaining point estimates of the signal parameters is concerned, the 
reported computational cost of the Bayesian algorithm appears to be significantly 
higher than either of the Frequentist methods. For example, 48 cores are 
used in \citet{2014PhRvD..90j4028T} to run a parallelized implementation of 
the \textit{MultiNest} algorithm \citep{2009CQGra..26u5003F} and the analysis is 
reported to typically take up to 45 minutes to complete at a network 
S/N $\rho_n=10$.  However, it should be noted that the Bayesian method also 
maps out the posterior probability distribution of parameters, which may provide 
useful information in an analysis. Interestingly, it has been demonstrated in the 
context of CMB analysis that a fitting procedure may be combined with PSO 
to map out the likelihood function locally around the point estimate \citep{2012PhRvD..85l3008P}. 
Thus, it may be possible to similarly extend MaxPhase (or WMJ1) to obtain information 
similar to that of a Bayesian method. This will lead to a corresponding increase in 
the computational cost of MaxPhase. 

\subsection{Effect of increasing the PTA size} \label{sec:morepsr}

As we noticed in the strong signal scenario, the maximization over the 
pulsar phases leaves behind a log-likelihood ratio that has strong secondary 
maxima, a feature that is absent if the pulsar phases are treated as 
intrinsic parameters. If these secondary maxima are comparable to the 
global maximum in value, they become attractors for stochastic search 
algorithms and reduce their effectiveness in locating the global maximum. 
With a decrease in signal-to-noise ratio, the probability of the locations 
of such secondary maxima becoming the global maximum increases. Both these 
effects worsen parameter estimation as we see in the moderate and weak 
signal cases.

This situation can be substentially improved by adding more pulsars in a PTA. 
Unlike the ground and space borne laser interferometers, adding more detectors 
(millisecond pulsars) in a PTA is technically easier and cheaper in terms of costs. 
Here, we demonstrate this by using the NANOGrav configuration \citep{2013ApJ...762...94D} 
which consists of 17 pulsars in the catalog.  We keep the network S/N $\rho_n$ 
the same for each scenario as in the analysis reported in 
Sec.~\ref{sec:strong}--\ref{sec:comp} with 9 pulsars. 
Accordingly, the overall amplitude $\zeta$ of the GW is scaled down. 
This implies that the signal amplitude for individual pulsars becomes significantly lower.

\begin{figure}[H]
\centerline{\includegraphics[width=1.4\textwidth, angle=0]{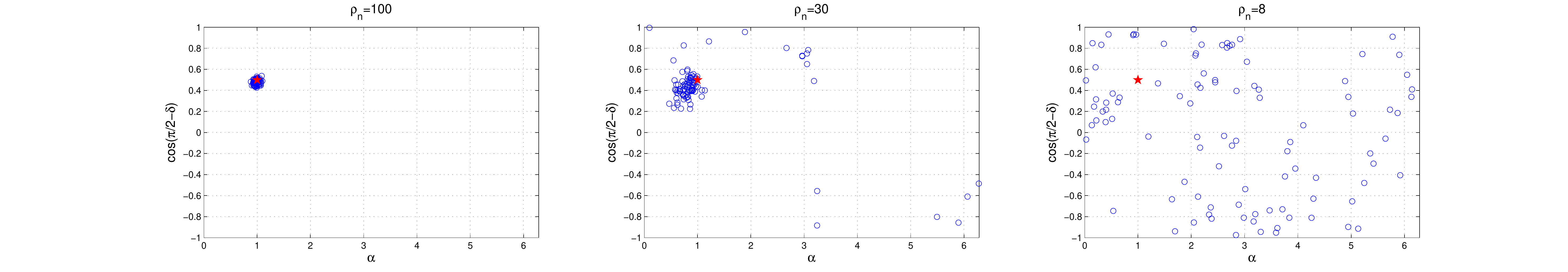}}
\caption{In each panel, blue circles show the estimated sky locations of the source, which are 
obtained from the MaxPhase algorithm for a PTA consisted of 17 pulsars. A red star marks the true 
location of the source used in the simulation. The x-axis represents Right Ascension and the 
y-axis represents the declination. The number of independent data realization is 100.}
\label{fig:pta17}
\end{figure}

Fig.~\ref{fig:pta17} presents the estimations of Right Ascension and declination of 
the GW source for 100 independent data realizations with $\rho_n=100$, 30 and 8 cases. 
Clearly, the scatter and the secondary maxima in the sky localization are effectively 
suppressed comparing to the ones in Fig.~\ref{fig:hist_snr3} and \ref{fig:hist_snr2} 
for the strong and moderate signal cases. For the weak signal case, although the localization 
is still inferior compared to WMJ1 and the Bayesian algorithm, the bias appearing 
in Fig.~\ref{fig:hist_snr1} is gone and the distribution becomes quite uniform.

\section{Summary and conclusions} \label{sec:summary}

Combined with WMJ1, this paper completes the first step in the program of 
implementing a purely Frequentist detection and parameter estimation approach 
for continuous wave GW signals using PTAs. There exists a dichotomy in how 
a GLRT can be implemented for this problem and this paper addresses the 
approach where pulsar phases are treated as extrinsic parameters that are 
maximized semi-analytically. Maximizing over pulsar phases is attractive 
compared to the alternative where they are treated as intrinsic parameters 
because the GLRT becomes scalable with the size of a PTA.  The maximization 
over the pulsar phases leaves behind a 7-dimensional numerical optimization 
problem irrespective of the number of pulsars in a PTA. We find that the 
latter problem is effectively handled using PSO, as was the case in WMJ1, 
without requiring much tuning. Computational costs of PTA data analysis 
methods will become especially important for analyzing the IPTA data set that 
includes about 50 pulsars.

The approach based on the analytical maximization over pulsar phases has the 
merit that it does not involve the type of constrained maximization that appeared 
in WMJ1. This greatly simplifies the implementation and boosts the computation  
speed of the method. However, our results indicate that the performance of 
the method is not as good as far as estimation of the source location and 
some of the other angular parameters is concerned. The increased errors appear 
to stem from secondary maxima. The fact that these secondary maxima disappear 
when the PTA size is increased, suggests that they are likely to be the 
result of not taking the ill-posedness 
of the GW network analysis problem -- well known in the context of ground based 
detector networks \citep{2005PhRvD..72l2002K, 2006CQGra..23.4799M} --  into account.

Mitigation of ill-posedness can be achieved by regularization of the inverse 
problem in some form \citep{1959SIAMR...1...38G,1977..book.....T,2006CQGra..23S.673R}. 
However, unlike ground based networks of large-scale detectors, we have the 
simple option in the case of PTAs to increase the number of independent 
detectors (i.e., pulsars).  In fact, the NANOGrav collaboration is adding 3-4 
new MSPs, discovered from the ongoing major pulsar surveys at Arecibo Observatory 
and Green Bank Telescope (e.g., PALFA and GBNCC), in the observation campaign 
every year. As known for the ground-based case, this should reduce the effect 
of ill-posedness. That this is so is shown explicitly by taking a PTA with a 
larger number of pulsars. However, although increasing the number of pulsars 
is an obvious way to mitigate the problem of ill-posedness, the results for 
the weak signal case --the realistic one for the current PTAs-- show that it cannot
be completely ignored and must be addressed properly. We leave a deeper look at the 
problem of ill-posedness and regularization to future work.

The results reported here were obtained under the following limitations. 
The simulated data was evenly sampled whereas real data will have irregular sampling. 
However, our method works entirely in the time domain, and no major changes are needed  
to accommodate irregularly sampled data.  In fact, if the irregularly sampled data 
have identically and independently distributed noise samples, no change in the 
algorithm is required. 
If, as some studies point out, the noise is not Gaussian or stationary, 
the actual covariance matrix for the given data will need to be modeled 
(or estimated) \citep{2015inPrepYWANG, 2015JPhCS.610a2019W}. 
Regarding non-Gaussianity, it is worth noting that \citet{2001PhRvD..63j2001F} 
shows that coherent techniques, such as MaxPhase and WMJ1, are generally 
robust against non-Gaussianity in the noise components.

The timing residuals for real data are obtained by fitting, using weighted least 
squares, a timing model to the data and subtracting it out. The timing model 
contains a set of parameters specific to the pulsar whose pulse arrival times 
are being fitted. The fitting procedure can affect the signal form as 
well as the statistics of the noise in the residual. 
When analyzing observational data, a common practice is to 
use the projection matrix \textbf{R} suggested by \citet{2013ApJ...762...94D}. 
A nice feature of \textbf{R} is that it only depends on the fitting model and the weighting 
matrix used, not the data itself. The influence of fitting can be easily taken into account by 
operating \textbf{R} on the timing residuals in the algorithm.

In constructing the GLRT, we assumed that the noise parameters are known \textit{a priori} 
or can be estimated independently of the GW analysis.  A more sophisticated approach would 
include the noise parameters as part of the estimation procedure. Since these additional 
parameters would be intrinsic in nature, directly including them in the GLRT would increase 
the search space dimensionality for PSO significantly. For example, the number of dimension 
increases from 7 to 52 for a PTA with 9 pulsars. Although such large dimensional optimization 
problems appear frequently in the PSO literature, it remains to be seen how the 
increase in dimensionality will pan out in the case of PTA data analysis. Some dimensional 
reduction scheme, of which fixing the noise model parameters \textit{a priori} is an extreme
example, will probably need to be implemented. 

Finally, our signal model does not include the ellipticity or the evolution of 
binary orbit during the period of observation. However, these modifications will 
only lead to a few more intrinsic parameters that are specific to the GW signal 
and not associated with the pulsars. A study of the GLRT approach for more 
sophisticated signal models will be carried out in future works.

\section{Acknowledgments}\label{ackng}

This work was supported by the National Science Foundation
under PIRE grant 0968296. The contribution of S.D.M. to
this paper is supported by NSF awards PHY-1205585 and
HRD-0734800.  Y.W. is supported by the 
National Science Fundation of China (NSFC) under grant NO. 11503007.
We are grateful to the members in the NANOGrav for helpful 
comments and discussions. 
We thank the anonymous referees for helpful comments. 
Y. W. acknowledges the hospitality of School of Physics at the University 
of Western Australia during his visiting, where part of this work has been done.

\bibliographystyle{apj}
\bibliography{pta4gw}

\end{document}